\documentclass[aps,prx,reprint,superscriptaddress]{revtex4-2}
\usepackage[T1]{fontenc}
\usepackage{hyperref}
\hypersetup{
  colorlinks   = true,
  urlcolor     = blue,
  linkcolor    = blue,    
  citecolor    = blue
}
\usepackage{physics,amsfonts,amsmath,graphicx,dcolumn,bm,array,color,amssymb,setspace,empheq,braket,multirow,cases,enumitem}
\usepackage[ruled,lined]{algorithm2e}

\begin{document}

\title{Atom-mediated deterministic generation and stitching of photonic graph states}

\author{Ziv Aqua}
\email{zivaqua@gmail.com}
\affiliation{AMOS and Chemical and Biological Physics Department, Weizmann Institute of Science, 76100 Rehovot, Israel}
\author{Barak Dayan}
\affiliation{AMOS and Chemical and Biological Physics Department, Weizmann Institute of Science, 76100 Rehovot, Israel}
\affiliation{Quantum Source Labs, Israel}

\date{\today}
             
\begin{abstract}
Highly-entangled multi-photon graph states are a crucial resource in photonic quantum computation and communication. Yet, the lack of photon-photon interactions makes the construction of such graph states especially challenging. Typically, these states are produced through probabilistic single-photon sources and linear-optics entangling operations that require indistinguishable photons. The resulting inefficiency of these methods necessitates a large overhead in the number of sources and operations, creating a major bottleneck in the photonic approach. Here, we show how harnessing single-atom-based photonic operations can enable deterministic generation of photonic graph states, while also lifting the requirement for photon indistinguishability. To this end, we introduce a multi-gate quantum node comprised of a single atom in a W-type level scheme coupled to an optical resonator. This configuration provides a versatile toolbox for generating graph states, allowing the operation of both the controlled-Z and SWAP photon-atom gates, as well as the deterministic generation of single photons. Furthermore, the ability to deterministically entangle photonic qubits enables expanding the generated state by stitching graphs from different devices. We investigate the implementation of this gate-based approach using $^{87}$Rb atoms and evaluate its performance through numerical simulations.
\end{abstract}

\maketitle
             
\section{Introduction}

Large-scale entanglement is at the core of quantum technologies, which hold the potential to surpass the capabilities of classical devices~\cite{preskill2018quantum}. At the forefront, fault-tolerant quantum computation relies on error-correcting codes in which logical quantum information is encoded in highly entangled states of many physical qubits~\cite{gottesman2010introduction}. As one of the most promising routes towards large-scale quantum computers, discrete-variable photonic quantum computing utilizes physical qubits encoded in single photons. Optical photonic qubits exhibit exceptionally low decoherence rates due to the absence of interaction between single photons and their weak interaction with the environment. Additionally, photons can be easily manipulated by linear optics to achieve high-fidelity single-qubit gates~\cite{carolan2015universal,harris2016large,alexander2024manufacturable}, and long optical fibers can essentially perform as nearly noise-free quantum memories~\cite{bombin2021interleaving}. However, the lack of photon-photon interaction poses a significant challenge in generating the necessary entanglement. In one-way~\cite{raussendorf2001one,raussendorf2003measurement,raussendorf2006fault,raussendorf2007topological,briegel2009measurement} and fusion-based~\cite{bartolucci2023fusion} quantum computing, which are particularly suited for the photonic platform~\cite{nielsen2004optical}, this challenge primarily lies in preparing fixed entangled resource states, referred to as cluster or graph states~\cite{briegel2001persistent,hein2004multiparty}. Additionally, photonic graph states serve as an entanglement resource for many more applications in photonic quantum information processing, including all-optical quantum repeaters for long-distance quantum communication~\cite{azuma2015all,azuma2023quantum} and quantum simulations~\cite{han2007scheme,lu2009demonstrating}.

The challenge of generating photonic graph states centers around two key tasks: producing single photons and performing photonic entangling operations. Traditionally, these tasks are carried out probabilistically, employing single-photon sources based on parametric photon pair generation~\cite{burnham1970observation,kwiat1999ultrabright,pittman2002single,sharping2006generation,reimer2014integrated,silverstone2014on}, and entangling operations that rely on multi-photon quantum interference in linear-optics circuits~\cite{browne2005resource,kok2007linear,varnava2008how,li2015resource,sparrow2018quantum,bartolucci2021creation}. Since quantum interference depends on the contribution of indistinguishable paths to the final state, ensuring high-fidelity entangling operations requires the sources to meet stringent conditions in terms of photon indistinguishability, presenting a major hurdle in the field. Notable demonstrations in this probabilistic approach include the observation of up to 12-photon graph states~\cite{bouwmeester1999observation,walther2005experimental,prevedel2007high,lu2007experimental,tao2012experimental,wang2016experimental,zhong2018photon} in bulk optics, and 4-photon graph states on a programmable photonic chip~\cite{adcock2019programmable,llewellyn2020chip}. A potential method to address the inefficiency associated with the generation process is to encode a number of qubits in multiple modes of a single photon, thereby reducing the number of photons required per graph state~\cite{vallone2007realization,chen2007experimental,vallone2008active,ceccarelli2009experimental,wang2018qubit,reimer2019high,erhard2020advances,vigliar2021error,lib2024resource}. Thus far, demonstrations in both these approaches have been restricted to schemes that produce graph states in an unheralded, postselected manner, where the generation of the state is ensured only upon measuring it. To achieve the on-demand operation necessary for quantum applications, it is essential to use heralded schemes that enable active multiplexing of these probabilistic processes. However, implementing such a procedure poses a considerable challenge, as it requires a substantial overhead in the number of heralded single-photon sources and entangling operations to guarantee the generation of graph states.

An alternative approach harnesses cavity-enhanced matter-based quantum emitters to deterministically generate single photons and mediate effective photon-photon interactions. The construction of photonic graph states in this approach has been explored in many theoretical works~\cite{schon2005sequential,lindner2009proposal,economou2010optically,buterakos2017deterministic,russo2019generation,tiurev2021fidelity,li2022photonic,hilaire2023neardeterministic,lobl2024loss}. Experimentally, small-scale states have been demonstrated with different types of quantum emitters, which serve either as single-photon sources at the input of linear-optics circuits~\cite{istrati2020sequential,li2020multiphoton,cao2024photonic,chen2024heralded,pont2024high}, or as memory qubits that distribute entanglement over a sequence of emitted photons~\cite{schwartz2016deterministic,yang2022sequential,thomas2022efficient,cogan2023deterministic,coste2023high,thomas2024fusion,meng2024deterministic}. However, realizing the large-scale photonic entanglement necessary for practical quantum computing requires a scalable design that integrates many quantum emitters. Crucially, the photonic states generated by these emitters must be compatible with the entangling operations that establish long-range quantum correlations. For instance, in fusion networks~\cite{bartolucci2023fusion}, these correlations are achieved by performing destructive entangling measurements on qubits from distinct graph states, which rely on multi-photon interference. Thus, to ensure the high-fidelity operation of these fusion measurements, it is essential to maintain high indistinguishability between photons generated by different emitters. This requirement is particularly challenging with solid-state emitters such as quantum dots, which, in contrast to atoms, are not inherently identical. Moreover, to retain the advantages of photonic quantum computing, the photon generation and entanglement processes must be both highly efficient and extremely rapid, such that the generated states can be stored in reasonable lengths of optical fibers without significant loss. Developing an efficient light-matter interface suitable for large-scale photonic architectures, which enables the rapid generation of high-fidelity photonic graph states, remains an outstanding challenge.

\begin{figure}
    \centering
    \includegraphics[width=\linewidth]{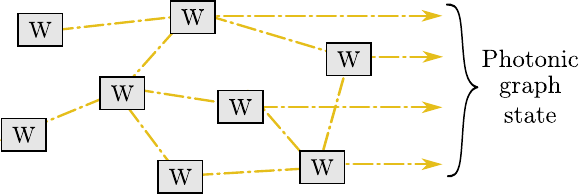}
    \caption{Conceptual depiction of a device comprising of optically interconnected W-type nodes, each with a single atom capable of performing photon-atom SWAP and CZ gates, as well as producing single photons. This construction provides a versatile platform for the deterministic generation of photonic graph states.}  
    \label{fig:W_interconnected}
\end{figure}

In this work, we propose a complete scheme for the deterministic and rapid generation of such photonic graph states based on cavity-mediated photon-atom quantum gates. We introduce a novel quantum node comprised of a single atom with a W-type level structure in an optical cavity. This node is capable of both generating single photons and executing two native photon-atom gates, namely SWAP and controlled-Z (CZ), using the same atomic qubit. The ability to perform these operations within the same apparatus enables a modular design, promoting the integration of such quantum nodes in photonic architectures, as depicted in Fig.~\ref{fig:W_interconnected}. Importantly, due to the inherently identical nature of atoms, photons emitted by one node are compatible with photon-atom gates in another node. Therefore, by combining the entangling photon-atom CZ gate, the ability to exchange atomic and photonic qubits via the SWAP gate, and readily available arbitrary photonic single-qubit gates, this gate-based approach provides a versatile framework for deterministic universal photonic quantum information processing. Specifically, it enables a variety of protocols for the deterministic generation of photonic graph states and introduces the unique capability of performing deterministic, nondestructive entanglement of photonic qubits from distinct graph states, in a process we term 'stitching'.

As previously noted, the rapid generation of graph states is crucial for photonic quantum computation, as it enables the use of long optical fibers as nearly decoherence-free quantum memories. This dictates timescales in the nanosecond regime, which is roughly 2-3 orders of magnitude faster than previous demonstrations relying on Raman transfers and repeated excitations of single atoms in free-space Fabry-P{\'e}rot cavities \cite{thomas2022efficient,thomas2024fusion}. Yet, this timescale is well within the typical duration of native photon-atom SWAP and CZ gates when implemented in compact, chip-based optical resonators. Additionally, the ability to integrate multiple atom-cavity nodes and execute simultaneous operations can enable even higher generation rates.

Furthermore, photon-atom gates are insensitive to the exact pulse shape of the photonic qubit, and accordingly can operate with sources that emit temporally impure photons, thereby lifting the strict requirement for indistinguishable photons. This insensitivity stems from the large separation of timescales between the long coherence lifetime of the atomic qubit, encoded in two ground states, and the fast light-matter interaction with the photonic qubit, dictated by the cavity-enhanced coupling rate with the excited state~\cite{rosenblum2011photon}. We show that high gate fidelities are achieved as long as the bandwidth of the photons is significantly smaller than the interaction bandwidth of the atom-cavity system, even when performed with photons in mixed spectro-temporal states. In other words, after the gate operation has ended, no knowledge of the transient dynamics of the interaction is imprinted on the atom or the outgoing photon.

This paper is structured as follows. First, in Sec.~\ref{sec:W}, we present a quantum node consisting of an atom with a W-type level structure coupled to an optical resonator, and specify its two operation modes for the photon-atom SWAP and CZ gates. In Sec.~\ref{sec:sps}, we describe a deterministic single-photon source based on the underlying mechanism of the SWAP gate. In Sec.~\ref{sec:gates} we establish the compatibility of this single-photon source with photon-atom gates. Following that, in Sec.~\ref{sec:graph}, we demonstrate how these operations can be harnessed to deterministically generate photonic graph states, using a module that produces 6-ring graph state as an example. We also describe the stitching process, which enables the deterministic entanglement of graph states generated by different modules. Subsequently, in Sec.~\ref{sec:feas}, we discuss the feasibility of this scheme, propose the implementation of the W-system in a $^{87}$Rb atom and asses its performance. Finally, in Sec.~\ref{sec:diss}, we summarize and offer concluding remarks. Throughout the paper, we present analytical calculations and numerical simulations detailed in the Appendices.

\section{Multi-gate node using a cavity-coupled atomic W-system}
\label{sec:W}

\begin{figure*}[t]
    \centering
    \includegraphics[width=\linewidth]{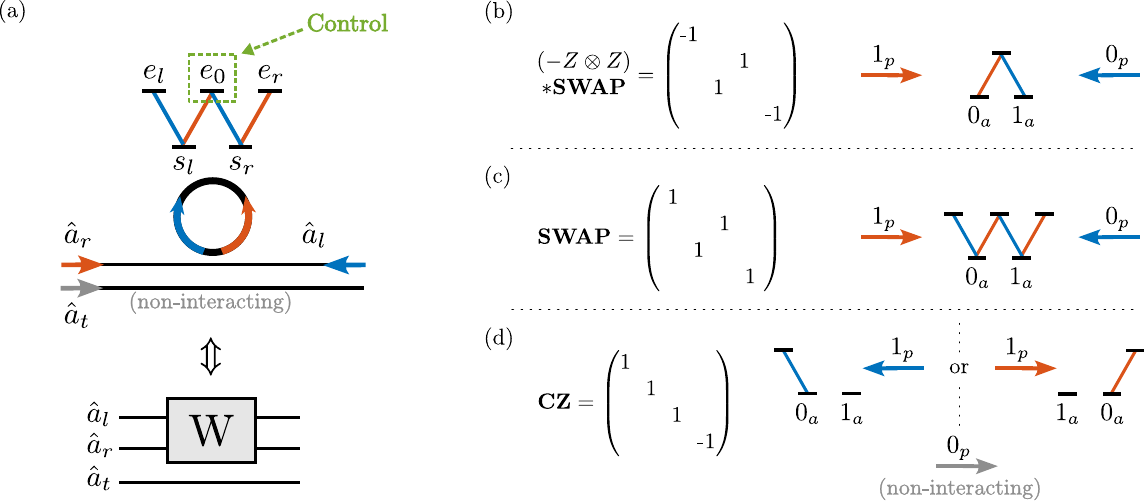}
    \caption{Multi-gate quantum node. (a) Schematic description of a single atom in a W-type level scheme coupled to two optical modes of a waveguide via a resonator; mode $\hat{a}_l$ couples to transitions $\ket{s_l}\leftrightarrow\ket{e_l}$ and  $\ket{s_r}\leftrightarrow\ket{e_0}$, while $\hat{a}_r$ is coupled to $\ket{s_r}\leftrightarrow\ket{e_r}$ and  $\ket{s_l}\leftrightarrow\ket{e_0}$. An additional guided mode, $\hat{a}_t$, does not interact with the resonator and the atom. At the bottom, we introduce a simplified representation of this W-system, with the convention that light propagates from left to right. In the following we refer to this diagram (without the mode labels) when discussing devices with multiple atom-cavity sites. (b)-(d) Configurations for different photon-atom gates. In all configurations, the two degenerate ground states of the atom encode the $\ket{0_a}$ and $\ket{1_a}$ states of the atomic qubit. (b) and (c) depict the operation of the SWAP gate in the $\Lambda$- and W-type level schemes, respectively. Note that the interaction with the additional excited states in the W-system counteracts and eliminates the $\pi$-phase of the empty cavity, which occurs in the $\Lambda$-system for the $\ket{0_a,0_p}$ and $\ket{1_a,1_p}$ input states. The photonic qubit is encoded in left- and right-propagating optical modes, coupled to distinct transitions in the atom, marked by blue and orange. These settings also support the SPRINT-based extraction of a single photon from a coherent pulse, as detailed in Sec.~\ref{sec:sps}. (d) The CZ gate operates when only one atomic transition is coupled to an optical mode, defining the state $\ket{1_p}$ of the photonic qubit, with the $\ket{0_p}$ state encoded in a non-interacting mode.}    
    \label{fig:W_system}
\end{figure*}

The photon-atom SWAP~\cite{koshino2010deterministic,bechler2018passive} and CZ~\cite{duan2004scalable,reiserer2014quantum} gates involve a cavity-mediated interaction of a single photon with a three-level atom. In both gates, the atom consists of a single excited level and two ground states, which encode the atomic qubit. However, each gate requires a different configuration of atom-cavity coupling. In the CZ gate, only one atomic transition is resonantly coupled to the cavity mode. Upon an incident single photon in that mode, the joint photon-atom state acquires a $\pi$-phase depending on the initial state of the atom. This conditional $\pi$-phase shift mechanism~\cite{reiserer2013nondestructive,tiecke2014nanophotonic} enables the operation of the CZ gate. On the other hand, the SWAP gate requires that both atomic transitions are exclusively coupled to two orthogonal modes of the cavity. This configuration leads to single-photon Raman interaction (SPRINT)~\cite{shomroni2014all,rosenblum2016extraction,rosenblum2017analysis}, where an incoming single photon effectively 'pushes' the atom to the dark state of the incoming mode. This serves as the underlying mechanism in the SWAP gate. To integrate both the conditional phase and the SPRINT mechanism within a single atom, additional atomic levels must be utilized.

We consider an atom in a W-type level scheme with two degenerate ground states, $s_{l}$ and $s_{r}$, and three degenerate excited states, $e_{l}$, $e_{0}$, and $e_{r}$. The atomic transitions are coupled to two orthogonal modes of a waveguide, $\hat{a}_{l}$ and $\hat{a}_{r}$, via an optical resonator, as illustrated in Fig.~\ref{fig:W_system}(a). An additional mode, $\hat{a}_t$, does not couple to the resonator and therefore has no impact on the atomic state dynamics. An external control field is employed to inhibit transitions to $e_0$ by interacting with atomic levels outside the W-system to either induce a light shift or optically dress this state. Disabling the coupling between the ground states through their common excited state, $e_0$, effectively divides the photon-atom interaction into two independent parts - left and right; specifically, mode $\hat{a}_\nu$ interacts solely with the $s_{\nu} \! \leftrightarrow \! e_\nu$ transition of the atom, where $\nu\in\{r,l\}$.

The cavity-mediated interaction can be described by the following Hamiltonian~\cite{gea2013space,gea2013photon},
\begin{alignat}{2}
\label{eq:Hamiltonian}
\mathcal{H}_{\text{int}} =  -i\hbar &\int&& d\omega\frac{\sqrt{\kappa/\pi}}{\kappa-i\omega} e^{-i(\omega+\delta_a)t} \\
& \times \Bigl(&&g_{r0}\hat{\sigma}^\dagger_{r0} \hat{a}_{l}(\omega) + g_{l0}\hat{\sigma}^\dagger_{l0}\hat{a}_{r}(\omega) \nonumber \\ 
& &&+ g_{ll}\hat{\sigma}^\dagger_{ll} \hat{a}_{l}(\omega) + g_{rr}\hat{\sigma}^\dagger_{rr}\hat{a}_{r}(\omega) \Bigr) +\text{h.c} \nonumber 
\end{alignat}
Here, $\hat{\sigma}_{\nu\mu} = \ket{s_\nu}\bra{e_\mu}$ represents the atomic lowering operator, where $\nu$ and $\mu$ label the ground and excited states of the atom, respectively. We denote the atom-cavity coupling strength of each transition by $g_\alpha$ for $\alpha\in\{ r0,l0,rr,ll\}$, where $2g_\alpha$ is the single-photon Rabi frequency. Additionally, we consider only symmetric transition strengths, i.e., $\abs{g_{r0}}=\abs{g_{l0}}$ and $\abs{g_{rr}}=\abs{g_{ll}}$. $\kappa$ represents the cavity amplitude decay rate, while $\delta_a$ and $\omega$ denote the detuning of the atomic transitions and the incident field, respectively, both relative to the cavity resonance frequency. Throughout this paper, the center frequency of the incoming optical field is resonant with the atomic transition, and unless stated otherwise, both are resonant with the cavity, i.e. $\delta_a=0$. In the interest of clarity, we simplify our discussion of the control field by assuming that, when activated, it effectively sets $g_{r0}\!=\!g_{l0}\!=\!0$. In section \ref{sec:feas}, where we explore the physical implementation of this scheme, we address this aspect in more detail. 

For simplicity, in what follows we assume ideal conditions for the operation of cavity-mediated photon-atom gates. First, we require a high cooperativity parameter, $C = \Gamma_\alpha/\gamma_\alpha \gg 1$, where $\Gamma_\alpha=g_\alpha^2/\kappa$ denotes the cavity-enhanced emission rate and $\gamma_\alpha$ represents the free-space spontaneous decay rate in each transition. Additionally, we assume that the intrinsic loss rate of the resonator, $\kappa_i$, is significantly smaller compared to the coupling rate between the resonator and the waveguide, $\kappa_e$. Under these circumstances, the effects of free-space spontaneous emission and intrinsic cavity losses are negligible, thus allowing the evolution of the system to be governed by the Hamiltonian specified in Eq. \ref{eq:Hamiltonian}. Consequently, in the ideal limit, we can conveniently set $\gamma=\kappa_i=0$. Finally, the bandwidth of the single photons used in the gates, denoted by $\sigma_\omega$, should satisfy the adiabaticity condition $\sigma_\omega \ll g_\alpha,\kappa,\Gamma_\alpha$. For brevity, whenever we do not include the spectro-temporal description of the incoming photon, it is assumed to satisfy this condition.
    
We investigate the evolution of the quantum state upon an incoming single photon and an atom initialized in one of its ground states. When the incident photon is in mode $\hat{a}_\nu$ and the atom is initialized in state $\ket{s_\nu}$, the photon effectively interacts with a two-level system in a cavity. The induced atomic dipole radiation in transition $\ket{s_{\nu}} \! \leftrightarrow \! \ket{e_\nu}$ destructively interferes with the intracavity field, rendering the atom-cavity system transparent~\cite{duan2004scalable,gea2013space}. As a result, the evolution of the photon-atom state is simply given by,  
\begin{align}
\label{eq:evo_2level}
    &a^\dagger_\nu\ket{0,s_{\nu}} \rightarrow a^\dagger_\nu\ket{0,s_{\nu}}
\end{align}
If the atom is instead initialized in the other ground state, $\ket{s_{\bar{\nu}}}$, the resulting output state depends on whether the control field is activated. When the control field is off, the atomic configuration results in SPRINT~\cite{koshino2010deterministic,gea2013photon,rosenblum2017analysis}; the induced dipole in the $\ket{s_{\bar{\nu}}} \! \leftrightarrow \! \ket{e_0}$ transition leads to destructive interference of the output in mode $\hat{a}_\nu$, which 'forces' the atom to emit a photon in the orthogonal mode, $\hat{a}_{\bar{\nu}}$, and toggle to $\ket{s_{\nu}}$, as described by Eq. \ref{eq:sprint}. However, when the control field is activated and the transition to $\ket{e_0}$ is suppressed, the incoming photon experiences a bare cavity and, accordingly, the state acquires a $\pi$-phase shift, as indicated in Eq. \ref{eq:condpi}.
\begin{subnumcases}{a^\dagger_{\nu}\ket{0,s_{\bar{\nu}}}\rightarrow}
   $$a^\dagger_{\bar{\nu}}\ket{0,s_{\nu}}$$ & : control off \label{eq:sprint}
   \\
   $$-a^\dagger_{\nu}\ket{0,s_{\bar{\nu}}}$$ & : control on \label{eq:condpi}
\end{subnumcases}
Lastly, when the incident photon is in the non-interacting mode, $\hat{a}_t$, the evolution of the photon-atom state  is naturally trivial,
\begin{equation}
\label{eq:evo_trivial}
    a^\dagger_t\ket{0,s_\nu} \rightarrow a^\dagger_t\ket{0,s_\nu}
\end{equation}

The operations described in Eq. \ref{eq:evo_2level}-\ref{eq:evo_trivial} form the basis for the SWAP and CZ photon-atom gates. As described in Fig.~\ref{fig:W_system}(c), when the control field is off, a SPRINT-based SWAP gate is realized with the following encoding,
\begin{align}
\label{eq:encodSWAP}
        & \ket{0}_{a} \triangleq \ket{s_l} \; ; \;  \ket{1}_{a} \triangleq \ket{s_r} \\
        & \ket{0}_{p} \triangleq \hat{a}^\dagger_l\ket{0}  \; ; \; \ket{1}_{p} \triangleq \hat{a}^\dagger_r\ket{0} \nonumber
\end{align}
where mode $\hat{a}_t$ remains idle. Note that in the $\Lambda$-type configuration~\cite{rosenblum2017analysis}, where excited levels $\ket{e_l}$ and $\ket{e_r}$ are absent, the evolution of the state in Eq. \ref{eq:evo_2level} acquires a $\pi$-phase shift as the photon interacts with an empty cavity. This results in a modified SWAP operation with local Pauli-Z unitaries, as depicted in Fig.~\ref{fig:W_system}(b). 

On the other hand, when the control field is activated, the atomic configuration enables the implementation of a CZ gate based on the conditional $\pi$-phase shift, utilizing the interaction with either the left or the right transition, as illustrated in Fig.~\ref{fig:W_system}(d). Here, the qubits are defined as follows,
\begin{align}
\label{eq:encodCZ}
        & \ket{0}_{a} \triangleq \ket{s_\nu} \; ; \;  \ket{1}_{a} \triangleq \ket{s_{\bar{\nu}}} \\
        & \ket{0}_{p} \triangleq \hat{a}^\dagger_t\ket{0}  \; ; \; \ket{1}_{p} \triangleq \hat{a}^\dagger_\nu\ket{0} \nonumber
\end{align}
and mode $\hat{a}_{\bar{\nu}}$ remains unused. Note that the distinct coupling with the left and right transitions can be harnessed to construct a robust CZ gate by incorporating modes $\hat{a}_l$ and $\hat{a}_r$ in a Sagnac interferometer~\cite{nagib2024robust}.

Overall, the ability to execute both gates within the same atomic level scheme establishes this configuration as a versatile node for universal photonic quantum information processing; effective photon-photon CZ gates are facilitated by combining the photon-atom SWAP and CZ gates, while photonic single-qubit gates can be implemented by linear-optics elements. When integrated with a suitable on-demand single-photon source, this system allows to efficiently generate arbitrary photonic graph states.

\section{Deterministic extraction of single photons}
\label{sec:sps}

\begin{figure}
    \centering
   \includegraphics[width=\linewidth]{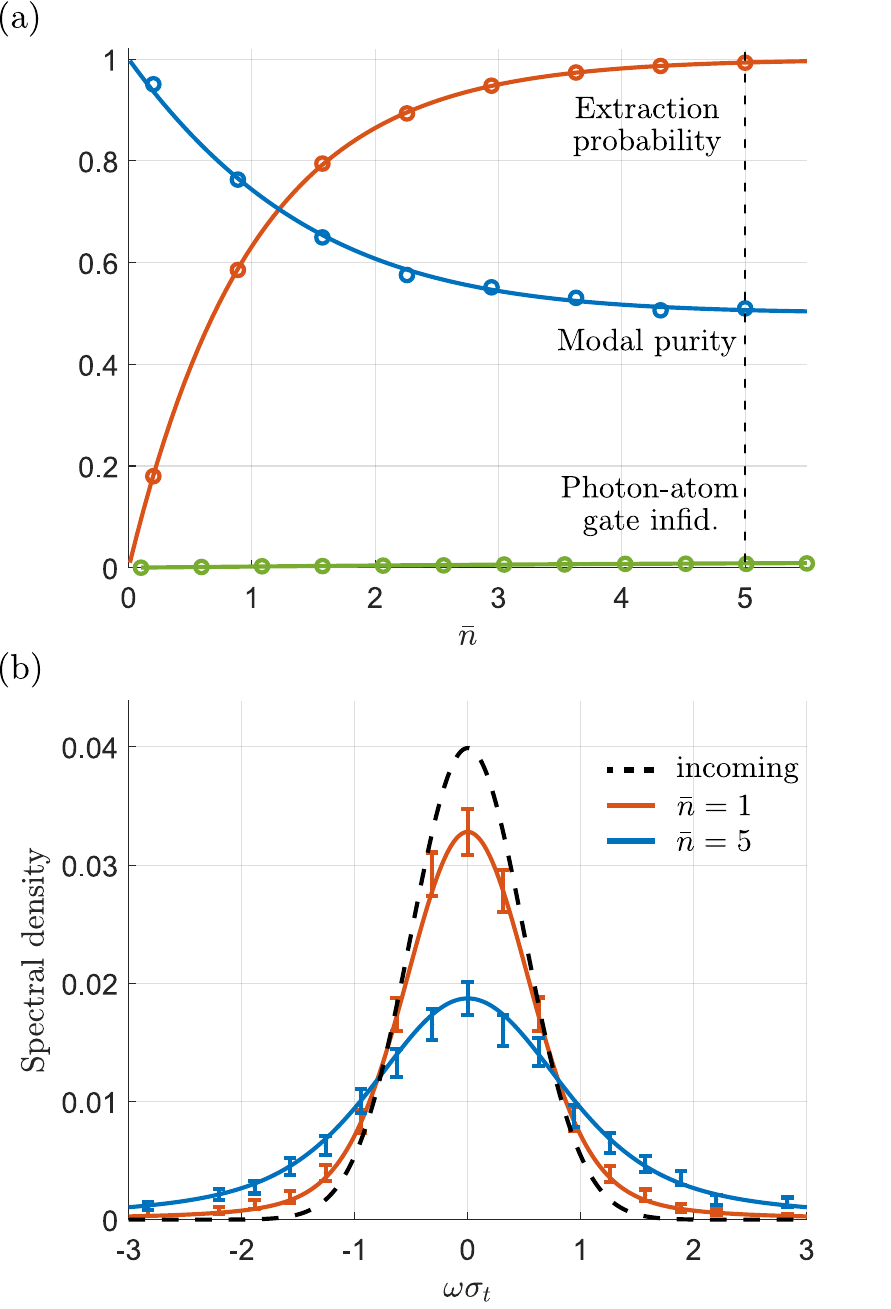}
    \caption{Single photon source based on extraction of single photons from coherent pulses by an atomic W-system. (a) Extraction success probability (orange), single-photon modal purity (blue), and infidelity of photon-atom gate performed with extracted photons (green) as a function of the average number photons in the incident coherent pulse, denoted by $\bar{n}$. Note that for $\bar{n}=5$, single photons are extracted practically deterministically, and even though the modal impurity approaches $1/2$, these photons can still be used to perform high-fidelity photon-atom gates (see Sec. \ref{sec:gates} for a detailed analysis). Solid lines represent analytical calculations, while circles denote numerical simulations (error bars are smaller than markers). (b) The spectral density of extracted photons for $n=1,5$, shown in orange and blue, respectively. Solid lines represent analytical calculations and error bars denote simulation results. The dashed line indicates the spectral density of the incoming pulse. Results in (a) and (b) are based on a Gaussian coherent pulse with $\sigma_t=50ns$ and an ideal source with $\kappa\sigma_t/2\pi=\Gamma\sigma_t/2 \pi=10^2$.}
    \label{fig:sps}
\end{figure}

The W-system can function as a single-photon source as 
SPRINT enables to deterministically extract a single photon from a classical coherent pulse~\cite{gea2013photon,rosenblum2016extraction}. When a multiple-photon pulse in mode $\hat{a}_r$ is incident upon a W-system initialized in $\ket{s_l}$, the first photon is extracted to mode $\hat{a}_l$ and the atom is toggled to $\ket{s_r}$, as outlined in Eq. \ref{eq:sprint}. Consequently, the remaining photons in the pulse are transmitted in mode $\hat{a}_r$. Ideally, failure can only occur due to the absence of a photon in the incoming pulse. Therefore, as the average number of photons in the pulse, $\bar{n}$, increases, the extraction success probability exponentially approaches unity (see Fig.~\ref{fig:sps}(a)). Following a successful extraction, a subsequent single photon can be extracted to mode $\hat{a}_r$ by sending a coherent pulse in mode $\hat{a}_l$, which also resets the atom to $\ket{s_l}$. This process can be repeated to generate a stream of single photons alternating between modes $\hat{a}_l$ and $\hat{a}_r$.  

Under ideal conditions, the final state of the system following a successful extraction process that started with the atom in $\ket{s_{\bar{\nu}}}$ is given by~\cite{gea2013photon},
\begin{align}
    \ket{\psi} = -e^{-\bar{n}/2} \sqrt{\bar{n}} &\int_{-\infty}^{\infty} dt f(t) a^\dagger_{\bar{\nu}}(t) \label{eq:extract} \\
    &\times \text{exp}(\sqrt{\bar{n}} \int_{t}^{\infty} dt' f(t') a^\dagger_\nu(t') ) \ket{0} \otimes \ket{s_\nu} \nonumber
\end{align}
where $f(t)$ represents the temporal wavepacket of the incoming pulse. Note that the single-photon nonlinearity introduced by the atom-cavity system ensures that the extracted field in mode $\hat{a}_{\bar{\nu}}$ does not contain higher Fock components, resulting in ideal number purity for this source. On the other hand, as indicated by the lower limit of the second integral in Eq. \ref{eq:extract}, the extracted photon and the transmitted field are time-entangled; only after the atom emits a photon in mode $\hat{a}_{\bar{\nu}}$ and becomes transparent to the incoming field can a detection event in mode $\hat{a}_\nu$ occur. Consequently, disregarding the temporal information in the transmitted field results in a single photon in a mixed state of temporal modes, with a modal purity approaching $1/2$ as $\bar{n}\rightarrow \infty$~\cite{gorshkov2013dissipative} (see Fig.~\ref{fig:sps}(a)). 

This fact seems to suggest that extracted photons may be unsuitable for linear-optics gates, which rely on interference of indistinguishable photons. However, a recent study shows that perfect two-photon interference can be recovered from a pair of extraction sources by employing a temporal quantum eraser (TQE)~\cite{aqua2024temporal}. This technique allows extracted photons to be used in two-photon linear-optics gates, making them compatible with fusion networks. In the TQE scheme, the incoming modes of two extraction sources are integrated within a balanced Mach-Zehnder interferometer. During the extraction process, the parity of detected photons in the dark port projects the extracted photons onto a two-photon state with a well-defined exchange symmetry, ensuring perfect two-photon interference. In Sec. \ref{sec:feas}, we investigate the feasibility of the TQE in a non-ideal configuration.

The more straight-forward approach we explore here is the use of photon-atom gates, which are insensitive to the temporal impurity of the photons. Still, the bandwidth of the extracted photons does have an effect on the performance of photon-atom gates in non-ideal atom-cavity systems. For a given incoming pulse, the probability of the first photon to arrive earlier within the pulse increases with $\bar{n}$. Therefore, the extracted photon pulse duration scales as $\sim1/\sqrt{\text{log}(\bar{n})}$~\cite{gorshkov2013dissipative}, leading to a broadening of its power spectrum (see Fig.~\ref{fig:sps}(b)). This effect can be countered by using longer coherent pulses at the input, at the price of a lower extraction repetition rate, which can then be mitigated by multiplexing several sources. In the following section, we examine the compatibility of this source with photon-atom gates in the W-system.

We note that alternative methods are available for producing pure single photons from atom-cavity systems. One such method involves fast excitation of the atom, followed by photon emission through the cavity mode~\cite{bochmann2008fast}. Yet, in this process, the bandwidth of the photon is completely determined by the spectrum of the atom-cavity system, making it unsuitable for high-fidelity photon-atom gates that require the adiabaticity condition discussed in Sec.~\ref{sec:W}. A more versatile method is vacuum-induced stimulated Raman adiabatic passage (vSTIRAP), which allows accurate control over the temporal envelope of the generated photon~\cite{morin2019deterministic}. However, this method requires resetting the atomic state after each photon generation, for example, through Raman population transfer. Thus, beyond the need for multiple precisely tuned external control fields, this method also introduces an additional time overhead.

\section{Photon-atom gates with extracted photons}
\label{sec:gates}
Photon-atom gates, particularly SWAP and CZ, can achieve near-ideal operation even when employing a single photon in a mixed state of temporal, and hence spectral modes, defined as follows,
\begin{equation}
    \rho = \iint d \omega_1 \omega_2 D(\omega_1,\omega_2) a^\dagger_{\nu}(\omega_1)\ket{0} \bra{0} a_{\nu}(\omega_2)
    \label{eq:mixedph}
\end{equation}
As we show below, this stems from the fact that the fidelity of these processes depends only on the power spectrum of the incoming photon, ${D(\omega,\omega)}$, rather than its entire spectral description, ${D(\omega_1,\omega_2)}$. Intuitively, in an ideal closed quantum system, the outgoing photon must 'carry' the information of the coherences, ${D(\omega_1,\omega_2)}$, making the final state of the atom insensitive to these off-diagonal terms. In other words, a measurement of the final atomic state cannot provide us with information on the spectral coherences of the photon. Therefore, extracted photons in mixed temporal states can be utilized reliably in photon-atom gates. 

In order to demonstrate this, we outline the derivation of the fidelity of SPRINT, which serves as the underlying mechanism of the SWAP gate, with an incident photon as defined in Eq. \ref{eq:mixedph}. A detailed derivation of both SPRINT and the conditional $\pi$-phase shift, which forms the basis of the CZ gate, are included in Appendix \ref{app:dirtygate}. We first examine a single photon in a pure state of mode $\hat{a}_r$ incident upon an atom initialized in $\ket{s_l}$. The final state of the system, following its evolution under the Hamiltonian in Eq. \ref{eq:Hamiltonian}, assuming $g_{r0}\!=\!g_{l0}\!=\!g$, is expressed by,
\begin{alignat}{2}
    \label{eq:SPRINT_evo}
    \ket{\psi_f}=&\int d\omega \tilde{f}(\omega) \Bigr[&& J(\omega) a^\dagger_l(\omega) \ket{0,s_r} \\ 
    & && + \Bigl(J(\omega) - \frac{\kappa + i\omega}{\kappa - i\omega} \Bigr) a^\dagger_r(\omega) \ket{0,s_l} \Bigl] \nonumber
\end{alignat}
Here, $\tilde{f}(\omega)$ is the  spectral amplitude profile of the incoming photon, and $J(\omega)$ is the spectral response of the atom-cavity the system, which depends on its parameters - $\kappa,g,\delta_a$  (see Appendix \ref{app:mou_sprint}).

When considering an incident photon in a mixed state, we can diagonalize the density matrix in Eq. \ref{eq:mixedph}, representing it as a statistical ensemble of pure single-photon states. Each of these states is evolved separately according to Eq. \ref{eq:SPRINT_evo}, and their mixture yields the final state of the system. The fidelity of the process is defined as the overlap the final state with the ideal state, where the atom is in $\ket{s_r}$ and the output photon in mode $\hat{a}_l$ (Eq. \ref{eq:sprint}). This fidelity is given by,
\begin{align}
    \label{eq:sprint_fid}
    \mathcal{F}_{sprint} = \int d\omega D(\omega,\omega) \abs{J(\omega)}^2 
\end{align}
which solely depends on the spectral bandwidth of the incoming photon. To ensure the high-fidelity operation of the gate, this bandwidth needs to be significantly narrower than the interaction bandwidth of the atom-cavity system, as defined by its physical parameters. A similar result arises when analyzing the fidelity of the condition phase mechanism (see Appendix~\ref{app:condpi}).

\begin{figure}
    \centering
    \includegraphics[width=\linewidth]{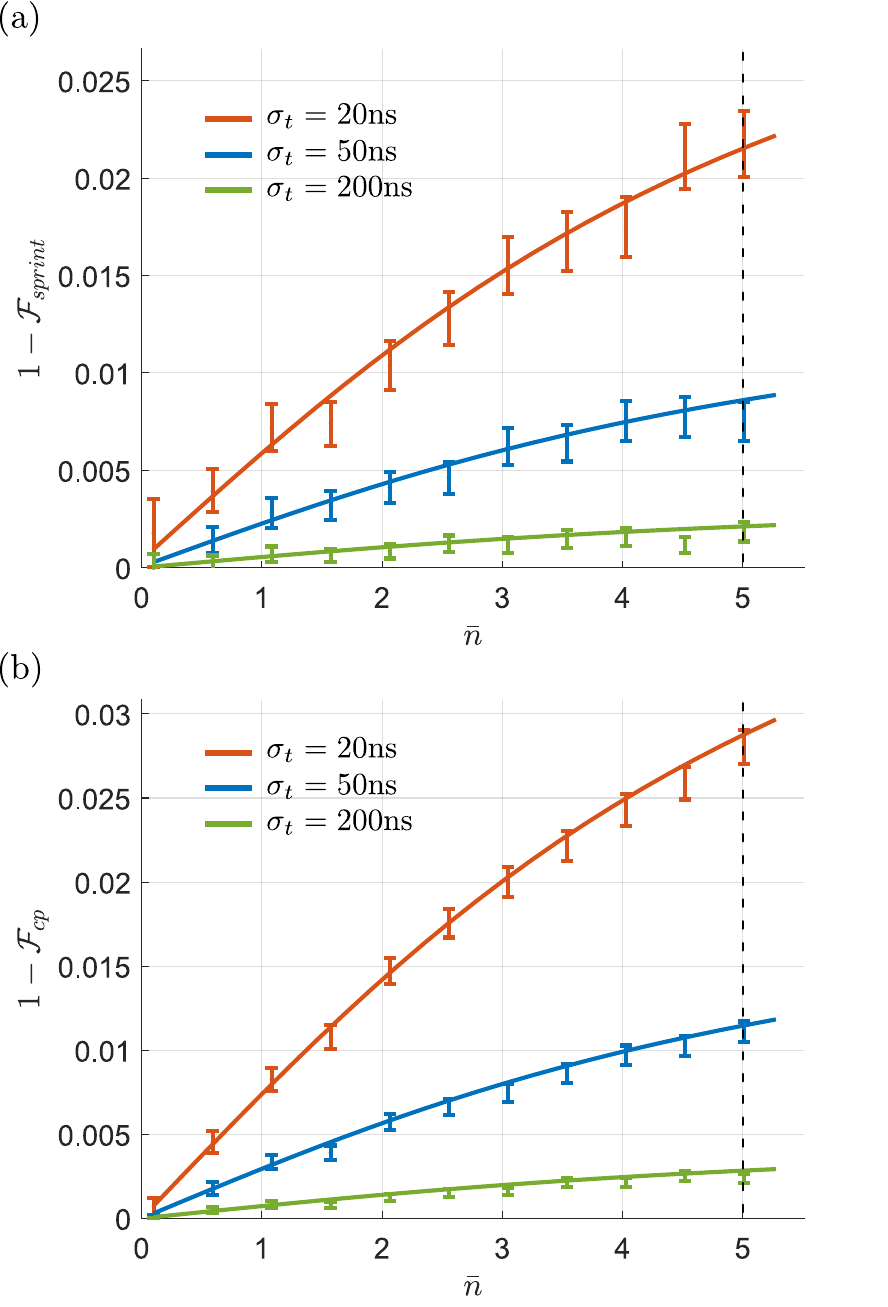}
    \caption{Infidelities in photon-atom gates with extracted photons. (a) and (b) shows the infidelity of SPRINT and the conditional $\pi$-phase shift, respectively. These operations are performed in a limited-bandwidth W-system with $g_\alpha=\kappa=2\pi\times100\text{MHz}$ using single photons extracted from coherent pulses with durations $\sigma_t=20,50,200\text{ns}$ (shown in orange, blue, and green, respectively). Solid lines represent analytical calculations, while error bars denote results from numerical simulations. Note that while the infidelity increases with the average photon number $\bar{n}$, this effect can be mitigated by employing a longer input coherent pulse. Specifically, with $\bar{n}=5$, corresponding to deterministic extraction, infidelities lower than 1\% can be achieved for the system parameters chosen.}    
    \label{fig:dirtygates}
\end{figure}

With this understanding, we can examine the fidelity of photon-atom gates with extracted photons. As previously discussed, the power spectrum of these photons broadens as the average number of photons in the initial coherent pulse increases, leading to gate infidelity in a limited-bandwidth atom-cavity system. To investigate this effect, we use an ideal W-system as a source unit that extracts a single photon from a coherent pulse with a Gaussian temporal envelope characterized by a standard deviation $\sigma_t$. This photon is directed into mode $\hat{a}_r$ of a different W-system, referred to as the gate unit, with a limited bandwidth set by $g_\alpha=\kappa=2\pi\times100\text{MHz}$. 

We consider two configurations for the gate unit; firstly, with the control field off, we examine the operation of SPRINT by preparing the atom in state $\ket{s_l}$. Secondly, we analyze the conditional phase mechanism by initializing the atom in $\ket{+}=(\ket{s_l}+\ket{s_r})/\sqrt{2}$ and activating the control field, which sets $g_{r0}\!=\!g_{l0}\!=\!0$. In the desired operation, the atomic state undergoes a phase flip, resulting in the orthogonal superposition state $\ket{-}=(\ket{s_l}-\ket{s_r})/\sqrt{2}$. Figure~\ref{fig:dirtygates} presents the infidelity in both processes resulting from the broadening of the extracted photon spectrum. As mentioned in the previous section, for given the bandwidth of the atom-cavity system, we can reduce the resulting infidelity by extracting single photons from longer coherent pulses.

Note that the conditional phase shift and SPRINT are the active parts of the CZ and SWAP gates, respectively; therefore, they correspond to the most demanding aspects of the photon-atom gates. Specifically, in the CZ gate, whenever the input photonic qubit is in $\ket{0}_p$ (Eq. \ref{eq:encodCZ}), the evolution of the state is trivial, and hence the desired gate operation is expected to occur with a unit fidelity. Similarly, in the SWAP gate, the operation for inputs $\ket{0_p0_a}$ and $\ket{1_p1_a}$ (Eq. \ref{eq:encodSWAP}) does not involve SPRINT, leading to ideal gate performance. Therefore, the state infidelity values in Fig.~\ref{fig:dirtygates} represent upper bounds to the gate infidelity of SWAP and CZ. 

\section{Deterministic generation of photonic graph states}
\label{sec:graph}

Graph states are a class of entangled states that can be associated with a graph consisting of vertices, representing qubits, and edges, indicating pairwise entanglement. These states can be constructed by initializing each of the qubits in $\ket{+}=(\ket{0}+\ket{1})/\sqrt{2}$ and performing a CZ gate between qubits that share an edge. A few interconnected W-type atom-cavity sites can be used to deterministically generate small-scale photonic graph states. The capabilities of W-systems, as introduced in the previous sections, provide a versatile toolkit for devising protocols aimed at producing such states; employing the CZ gate to generate photon-atom entanglement, mapping between atomic and photonic qubits by applying the SWAP gate, and generating single photons compatible with these processes through SPRINT-based extraction. For consistency, we henceforth define the atomic qubit as described in Eq. \ref{eq:encodSWAP},
\begin{align}
\label{eq:encodSWAP2}
        & \ket{0}_{a} \triangleq \ket{s_l} \; ; \;  \ket{1}_{a} \triangleq \ket{s_r}
\end{align}

\subsection{Dual-CZ gate}

\begin{figure}
    \centering
    \includegraphics[width=\linewidth]{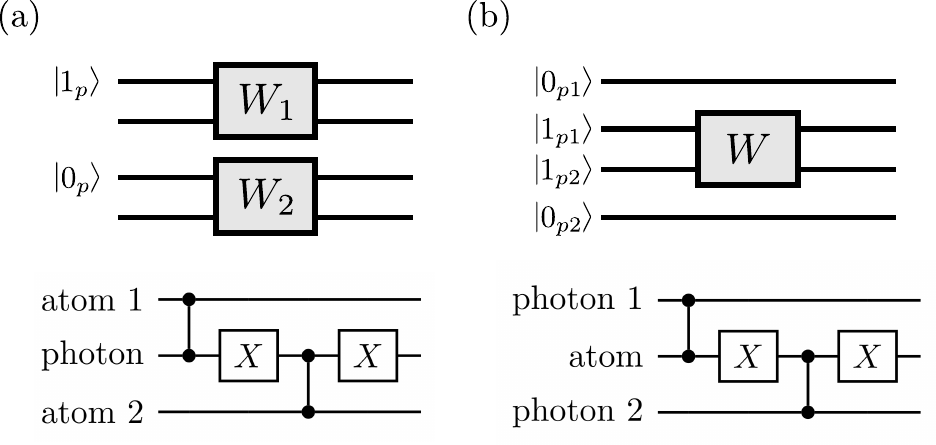}
    \caption{Schematic description and the equivalent quantum circuit for two schemes of a dual-CZ gate (DCZ) between three qubits: (a) a single photonic qubit and two atomic qubits in separate W-systems, or (b) a single atomic qubit and two photonic qubits.Note that the Pauli-X operations in the equivalent circuits arise from our definition of the qubits and do not correspond to physical operations.} 
    \label{fig:CZ}
\end{figure}

Naturally, achieving the desired output state involves a tradeoff between the duration of a protocol and the number of atom-cavity sites needed to execute it, considering that certain photon-atom operations can take place simultaneously. However, two key observations regarding the implementation of the CZ gate in the W-system allow us to utilize resources more efficiently. First, in the CZ gate, the state $\ket{0_p}$ corresponds to a photon in a non-interacting mode $\hat{a}_{t}$ (Eq. \ref{eq:encodCZ}). Instead, this mode can be directed to a different W-system, as depicted in Fig.~\ref{fig:CZ}(a), enabling us to simultaneously execute two CZ gates between a single photonic qubit and two atomic qubits in separate W-systems, denoted $W_1$ and $W_2$. We refer to this operation as a dual-CZ gate (DCZ), where the CZ gate with $W_1$ proceeds as usual, while the gate operation with $W_2$ effectively includes Pauli-X unitaries on the photonic qubit as mode $\ket{0_p}$, rather than $\ket{1_p}$, interacts with this atom. It is important to emphasize that the DCZ gate does not require additional Pauli-X physical operations; these are simply a consequence of our definition of the qubit states.

Second, as previously stated, when the control field is activated, the W-system configuration consists of two separate photon-atom cavity-enhanced interactions, where mode $\hat{a}_l$ ($\hat{a}_r$) couples exclusively to transition $\ket{s_{l}} \! \leftrightarrow \! \ket{e_l}$ ($\ket{s_{r}} \! \leftrightarrow \! \ket{e_r}$). This configuration enables to perform a DCZ gate with a single atomic qubit and two photonic qubits, as described in Fig.~\ref{fig:CZ}(b). In this setup, the $\ket{1}$ state of the first (second) photonic qubit is encoded in mode $\hat{a}_l$ ($\hat{a}_r$) of the W-system, whereas the $\ket{0}$ state remains uncoupled to the atom-cavity system. Once again, due to the definitions of the qubits, the CZ gate with the second photonic qubit effectively involves Pauli-X gates on the atomic qubit.

\subsection{Example: 6-ring graph state}

\begin{figure}
    \centering
    \includegraphics[width=\linewidth]{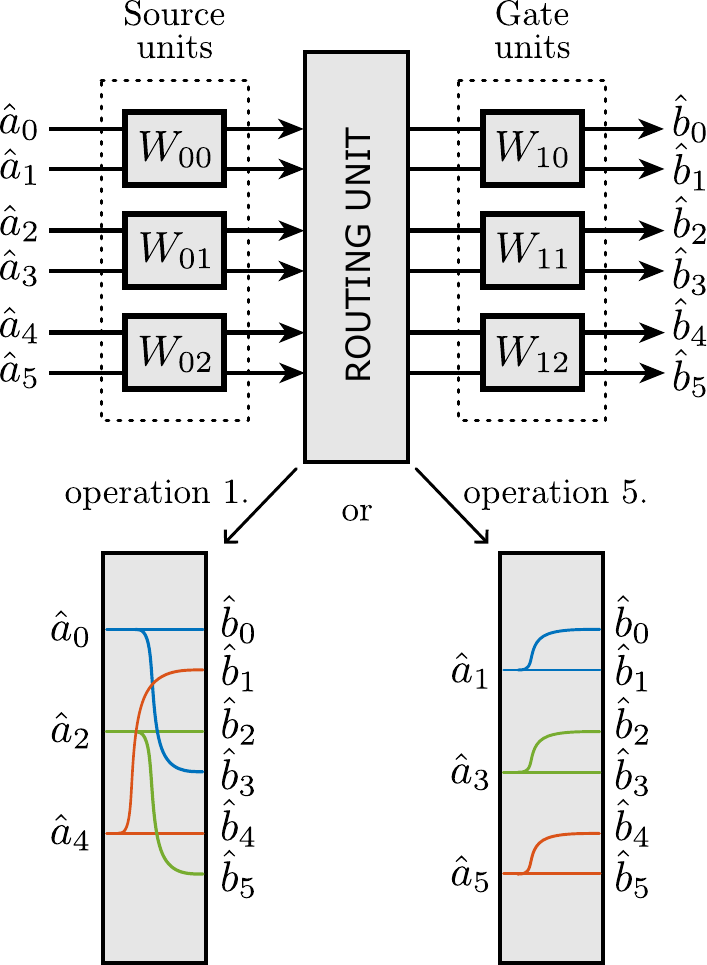}
    \caption{Schematic description of a device for the generation of 6-ring graph states. The setup is comprised of six W-type atom-cavity sites interconnected through a routing unit that can be set in two configurations corresponding to operations 1 and 5 of Algorithm \ref{alg:6ring}.}    
    \label{fig:6ring_setup}
\end{figure}

To demonstrate the concepts discussed so far, we present an efficient scheme for generating 6-ring graph states utilizing six W-type atom-cavity sites. We consider the configuration shown in Fig.~\ref{fig:6ring_setup}, comprising of three source units, labeled $W_{0n}$, and three gate units, denoted $W_{1n}$, where $n\in\{0,1,2 \}$. These units are interconnected via a routing unit with input and output modes denoted by $\hat{a}_{0-5}$ and $\hat{b}_{0-5}$, respectively. Initially, each source atom $W_{0n}$ is prepared in its left ground state, ready to extract a single photon in mode $\hat{a}_{2n}$, and each gate atom is prepared in its $\ket{+}_a$ state, functioning as a intermediate atomic node in the graph state. The generation sequence is then initiated, and can be repeated as required. 

\begin{figure}
    \centering
    \includegraphics[width=\linewidth]{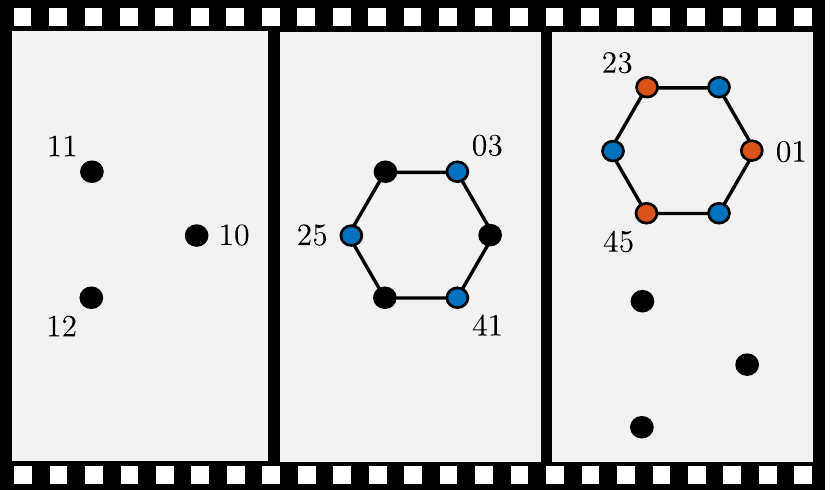}
    \caption{The three-step evolution of the generation process in Algorithm \ref{alg:6ring}; initialization of the atomic qubits, generation of photon-atom entanglement with DCZ gates, and mapping of atomic qubits to output photonic qubits using SWAP. In the diagram, black vertices represent the atomic qubits in the gate units, with labels $1n$ indicating each W-system. Blue and orange vertices labeled by $ij$ represent output photonic qubits, each defined by modes $\hat{b}_i$ and $\hat{b}_j$. The SWAP operation in the last step allows for the re-initialization of atomic qubits, enabling the repetition of the last two steps as necessary.}    
    \label{fig:6ring_steps}
\end{figure}

At the first step, we produce an intermediate photon-atom 6-ring graph state, comprised of the atomic qubits in $W_{1n}$ and the photons generated from $W_{0n}$ (see Fig.~\ref{fig:6ring_steps}). By sending coherent pulses in modes $\hat{a}_{2n\!+\!1}$, single photons are extracted in modes $\hat{a}_{2n}$ and source atoms are toggled to their right ground state. The routing unit is configured to direct modes $\hat{a}_{2n}$ into a superposition $(\hat{b}_{2n} + \hat{b}_{2n\!+\!3})/\sqrt{2}$, where mode index addition is modulo 6. This results in three photonic qubits at the inputs of the gate units, each initialized in the state $\ket{+}_n$ defined by modes $\hat{b}_{2n}$ and $\hat{b}_{2n\!+\!3})$. The external control field is activated at the gate units $W_{1n}$ to set them in the CZ operation mode and three simultaneous DCZ gates are implemented; each photonic qubit $\ket{+}_n$ undergoes two CZ gates with the atomic qubits in sites $W_{1(2n)}$ and $W_{1(2n\!+\!1)}$, where atom index addition modulo 3 is understood.

After the coherent pulses pass through the source units, we initiate the second step, aiming to map the atomic qubits of the intermediate graph state into photonic qubits. Since the atoms in $W_{0n}$ transitioned to the right ground states at the first extraction event, coherent pulses in modes $\hat{a}_{2n}$ enable us to extract single photons in modes $\hat{a}_{2n\!+\!1}$ and reset the source atoms to their left ground states. We configure the routing unit to direct $\hat{a}_{2n\!+\!1}$ into a superposition $(\hat{b}_{2n} + \hat{b}_{2n\!+\!1})/\sqrt{2}$ and deactivate the control fields in $W_{1n}$, setting them in a SWAP operation mode. In this configuration, each photonic qubit, now defined in modes $\hat{b}_{2n}$ and $\hat{b}_{2n\!+\!1}$, is incident upon gate unit $W_{1n}$. Accordingly, three SWAP gates are applied, serving two purposes; mapping the atomic qubits into photonic qubits, thus completing a 6-ring photonic graph state, and resetting the atomic qubits in $W_{1n}$ to $\ket{+}$ for the next cycle of the generation scheme. Overall, the repetition rate of the protocol is dictated by the duration of the extraction pulses in its two steps. The sequence of operations for the protocol are summarized in Algorithm \ref{alg:6ring}.

\begin{algorithm}
\label{alg:6ring}
\caption{generation of 6-ring graphs}
\begin{enumerate}[leftmargin=0.1cm,start=0] 
    \item Initialization:\\
    $W_{0n}$ source atoms in left ground state\\
    $W_{1n}$ gate atoms in qubit state $\ket{+}$
    \item Set routing unit:\\
    $\hat{a}_{2n} \rightarrow (\hat{b}_{2n} + \hat{b}_{2n+3})/\sqrt{2}$ 
    \item Turn control field ON in gate units $W_{1n}$
    \item Send extraction pulses in modes $\hat{a}_{2n+1}$
    \item Wait pulse duration
    \item Set routing unit:\\
    $\hat{a}_{2n+1} \rightarrow (\hat{b}_{2n} + \hat{b}_{2n+1})/\sqrt{2}$
     \item Turn control field OFF in gate units $W_{1n}$
    \item Send extraction pulses in modes $\hat{a}_{2n}$
    \item Wait pulse duration
    \item Repeat operations 1-8
\end{enumerate}
\end{algorithm}

Note that the due to effective Pauli-X gates involved in the operation of the DCZ gates (see Fig.~\ref{fig:CZ}(a)-(b)), the output photonic state requires the application of a $Z^{\otimes 6}$ unitary in order to conform with the standard definition of a graph state. Implementing this Pauli-Z correction for each qubit can be easily achieved in photonic circuits using phase elements. However, this correction may not be necessary, as it simply introduces a Pauli transformation to the stabilizers of the graph state, which can be accounted for when employing this state as an entanglement resource.

\subsection{Large-scale entanglement}

\begin{figure*}[t]
    \centering
    \includegraphics[width=\linewidth]{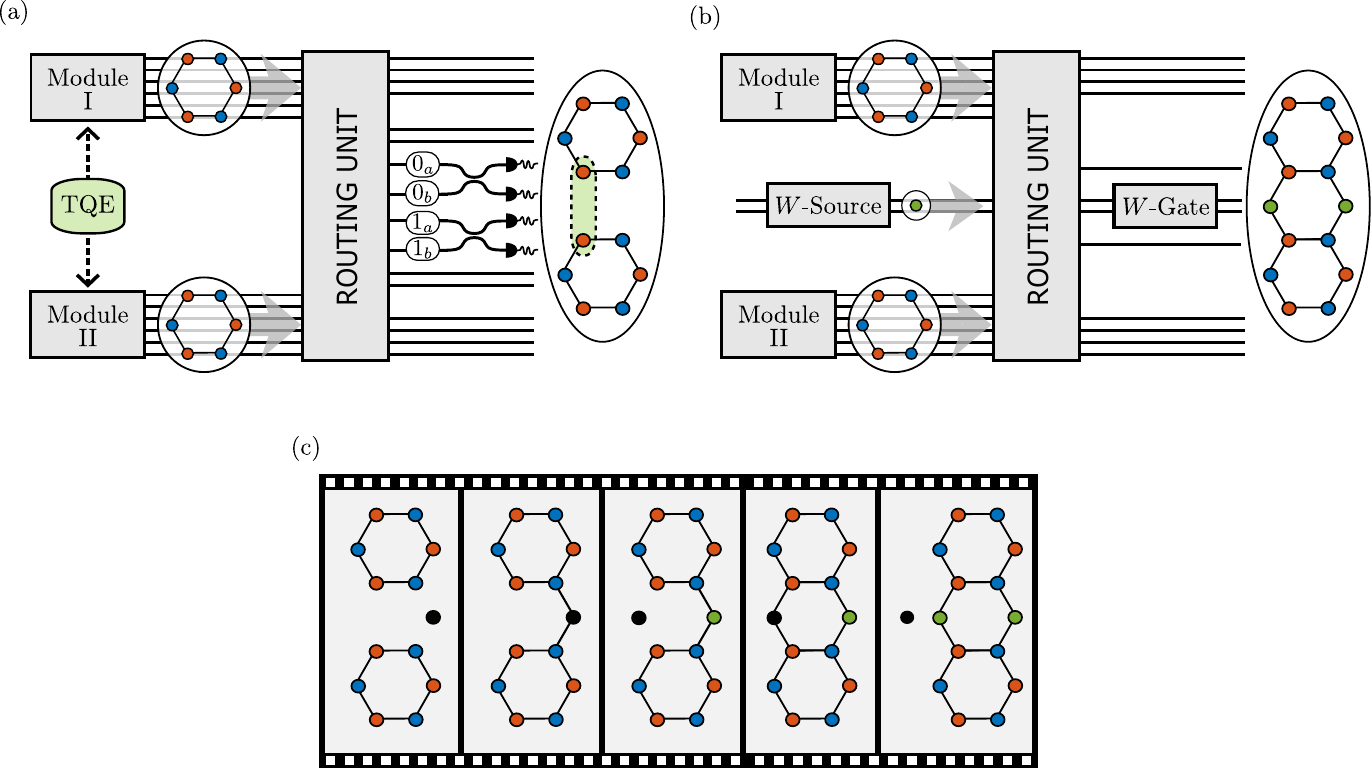}
    \caption{Entangling photonic graph states generated by different modules. (a)-(b) Schematic description of two possible approaches - fusion and stitching. (a) The probabilistic fusion gate, in its basic form, relies on a destructive measurement of two photonic qubits, following 50:50 beam-splitters that interfere the $\ket{0}$ and $\ket{1}$ qubit modes, respectively. Here, we show a fusion Upon specific detection patterns, the qubits are projected into an entangled state. Notably, extracted photons can be made compatible with these linear-optics gates via the TQE scheme. (b) The stitching method involves directing photonic qubits to a stitching module consisting of two W-systems: a source unit and a gate unit. Through photon-atom gates, this enables deterministic and nondestructive entanglement of photonic qubits. (c) Evolution of the stitching process for two 6-ring graph states. Initially, the atomic qubit (black) in the gate unit is prepared in the $\ket{+}$ state. Two photonic qubits (blue) are then routed to the gate unit to perform a DCZ gate. Next, the source unit emits a photon, which is directed to the gate unit to perform a SWAP gate, mapping the atomic qubit to a photon and resetting the atom to $\ket{+}$. This process is then repeated for the next pair of photonic qubits (orange).}    
    \label{fig:stitch}
\end{figure*}

A modular photonic architecture, where small-scale graph states are generated at each module, offers significant advantages by enabling efficient allocation of physical resources and better control over error propagation. In these architectures, entangling qubits from graph states generated by different modules is required both for increasing the size of small-scale graph states and for achieving the large-scale quantum correlations necessary for useful quantum computation. A critical aspect in this process is ensuring that the output photons from each module are compatible with the intended entangling operation.

Thus far, probabilistic fusion gates, which rely on destructive measurements of photons, have been the primary method considered for this task \cite{gimeno2015from,bartolucci2021creation,bartolucci2023fusion,paesani2023high,song2024encoded,chan2024tailoring}. Yet, the high-fidelity operation of these gates imposes stringent requirements on the indistinguishability of the photons involved, presenting a significant challenge, especially in architectures utilizing quantum emitters as photon sources. As described in \cite{aqua2024temporal}, pairs of extracted photons can be made compatible with two-photon linear-optics gates through the application of the TQE. Hence, employing the TQE scheme between source units in different modules enables entangling fusion measurements to connect graph states generated by these modules (see Fig.~\ref{fig:stitch}(a)).

Our gate-based approach, however, enables a deterministic and nondestructive method of entangling qubits from different graph states, in a process we term 'stitching'.
As a key feature of our approach, due to the inherently identical nature of atoms, a photon leaving one atom-cavity site can seamlessly interact with other atom-cavity sites. This capability allows photonic qubits from different graph states to be routed into a stitching module, where they can be deterministically entangled through photon-atom gates. As discussed in Sec.~\ref{sec:gates}, these gates are inherently robust to variations in the temporal mode of the photons, thus relaxing the demand for photon indistinguishability.

As an example, Fig.~\ref{fig:stitch}(b)-(c) illustrates the process of entangling two 6-ring graph states using a stitching module composed of two W-systems: one serves as a source unit and the other as a gate unit. The procedure begins with initializing the atomic qubit in the stitching gate unit in the $\ket{+}$ state. Next, two photonic qubits, each emitted from a different module, are directed to the gate unit to perform a DCZ gate. The stitching source unit then extracts an additional photon, which is routed to the gate unit to perform the SWAP gate, mapping the atomic qubit to a photonic qubit and resetting the atomic qubit in $\ket{+}$. This process can then be repeated for two subsequent photonic qubits emitted from the 6-ring modules.

\section{Feasibility}
\label{sec:feas}

The proposed setup for generating photonic graph states relies on trapping single atoms within low-loss, high-cooperativity optical resonators. Single-atom trapping has been achieved in a wide range optical cavities, with different cooling schemes developed to ensure a reliable photon-atom interface~\cite{reiserer2015cavity}. Achieving such an interface in scalable devices is essential to enable practical applications in quantum computing and communication. Recent demonstrations of potentially scalable systems include fiber Fabry-P{\'e}rot~\cite{gehr2010cavity,volz2011meas,gallego2018strong,brekenfeld2020quantum}, tapered nanofiber~\cite{kato2015strong,nayak2019real}, photonic crystal~\cite{thompson2013coupling,samutpraphoot2020strong,dordevic2021entanglement}, and whispering-gallery-mode~\cite{will2021coupling} resonators. In this context, trapping single atoms near photonic integrated circuits, compatible with state-of-the-art fabrication techniques, is highly desirable. However, this task is particularly challenging due to limited optical access caused by the typical dimensions of the photonic chip.  Nevertheless, recent advancements in chip design and trapping methods have shown significant promise~\cite{kim2019trap,chang2019microring,chang2020efficiently,zhou2023coupling}.

\begin{figure}
    \centering
    \includegraphics[width=\linewidth]{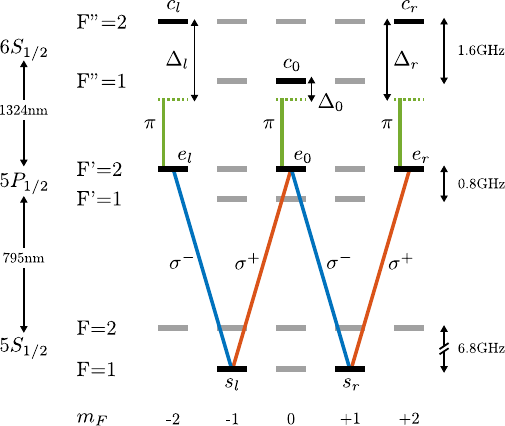}
    \caption{A possible physical realization of the multi-gate W-system in $^{87}$Rb. The W-system is comprised of two ground states within the $5S_{1/2}$ manifold and three excited states in $5P_{1/2}$, with the relevant $\sigma^{\pm}$-polarized transitions at $795$nm. We utilize a $\pi$-polarized external control field at $1324$nm to selectively inhibit transitions in the W-system by coupling each of the excited states to a higher excited state in $6S_{1/2}$.}    
    \label{fig:Rb}
\end{figure}

We investigate the implementation of the atomic W-type level scheme in the $D_1$ line of $^{87}$Rb ($5S_{1/2}\!\rightarrow\!5P_{1/2}$), as illustrated in Fig.~\ref{fig:Rb}. The states of the W-system are defined as follows,
\begin{align}
    &\ket{s_{l,r}} = \ket{F=1,m_F=-1,1} \\
    &\ket{e_{l,0,r}} = \ket{F'=2,m_F=-2,0,2} \nonumber
\end{align}
Here, the optical modes $\hat{a}_r$ and $\hat{a}_l$ couple to the $\sigma^+$ and $\sigma^-$ transitions, respectively. It is important to note that this configuration arises naturally in the atomic levels of $\ket{F=1}\!\rightarrow\!\ket{F'=2}$, and the optical coupling to the relevant $\sigma^{\pm}$ transitions is an inherent feature of counterpropagating transverse-magnetic modes in whispering-gallery-mode microresonators~\cite{junge2013strong}. We consider a $\pi$-polarized external control field, resonant with the $5P_{1/2}\rightarrow6S_{1/2}$ transition. This field couples each excited states in the W-system to a higher excited state in the $6S_{1/2}$ manifold, denoted as,
\begin{align}
    &\ket{c_{0}} = \ket{F''=1,m_F=0} \\
    &\ket{c_{l,r}} = \ket{F''=2,m_F=-2,2} \nonumber
\end{align}
The interaction Hamiltonian is given by,
\begin{equation}
    \mathcal{H}_\text{control} = \sum_{\mu \in \{ l,0,r\}} \Omega_{\mu} \sigma'_{\mu} e^{-i\Delta_\mu t} + \text{h.c.}
\end{equation}
where $\hat{\sigma}'_{\mu}=\ket{e_\mu}\bra{c_\mu}$ denotes the transition lowering operator and $\Omega_\mu$ represents the Rabi frequency of the external field in that transition. The detuning of the external field relative to the transitions is denoted by $\Delta_\mu$, where $\Delta_l = \Delta_r$ are related to $\Delta_0$ by the frequency difference between $F''=1$ and $F''=2$. When $\Delta_0 = 0$, the control field dresses the states $\ket{e_0}$ and $\ket{c_0}$, leading to electromagnetically induced transparency that inhibits atomic absorption from $\ket{s_{l,r}}$ to $\ket{e_0}$. Additionally, a light shift is introduced to $\ket{e_l}$ and $\ket{e_r}$ due to the interaction of the detuned field with transitions $\hat{\sigma}'_{l,r}$. Alternatively, when $\Delta_{l,r} \!=\! 0$, transitions $\ket{s_l}\!\leftrightarrow\!\ket{e_l}$ and $\ket{s_r}\!\leftrightarrow\!\ket{e_r}$ are suppressed, while $\ket{e_0}$ experiences a light shift. This effectively transform the atomic level scheme from a W- to a $\Lambda$-configuration. The extent of these effects is related to the ratio between $\Omega$ and $g_\alpha$, and in the suitable parameter regime, they can serve as a reliable tool for suppressing transitions in the W-system.

We conduct numerical simulations (see Appendix \ref{app:simulation}) to evaluate the performance of the photon-atom operations in this implementation with the following parameters. We utilize a resonator characterized by $(\kappa_{e},\kappa_i)/2\pi=(100,1)~\text{MHz}$, corresponding to an intrinsic quality factor of $\sim\!10^8$. The atom-cavity coupling strength in the cycling transition of the $D_2$ line of $^{87}$Rb is $g_{cyc}/2\pi = 200\text{MHz}$. The coupling strengths, $g_\alpha$, are derived from $g_{cyc}$ using the ratio between the dipole matrix elements of the relevant transitions, resulting in $(g_{ll},g_{rr},g_{l0},g_{r0})/2\pi \approx -(141,141,58,58) \text{MHz}$. Although values within this range for $g_{cyc}$ and $\kappa_i$ have thus far only been demonstrated in separate devices~\cite{gehr2010cavity,volz2011meas,thompson2013coupling,dordevic2021entanglement,samutpraphoot2020strong,jin2021hertz,ji2021methods}, we believe that realizing such a photon-atom interface is possible in the near future. The CZ configuration of the W-system is enabled by activating the control field with $(\Omega,\Delta_0)/2\pi = (100,0)\text{MHz}$. Additionally, in order to optimize photon-atom operations in the presence of multiple excited levels~\cite{rosenblum2017analysis}, the resonance frequency of the cavity is slightly detuned from the $F\!=\!1\!\rightarrow\!F'\!=\!2$ transition, with $\delta_a/2\pi=-6~\text{MHz}$.

We characterize the SWAP and CZ gates by employing pure single photons with a Gaussian temporal profile of $\sigma_t=25\text{ns}$, resonant with the $F\!=\!1\!\rightarrow\!F'\!=\!2$ transition. As previously discussed, the gate infidelity arising from utilizing photons in mixed states is determined only by the power spectrum of these photons relative to the atom-cavity bandwidth. Therefore, our simulation results also apply for extracted photons, as the duration of the initial coherent pulses in the extraction process can always be adjusted to ensure that the spectrum of the these photons is narrower than that of the pure photons used in the simulation. We perform quantum process tomography~\cite{chuang1997prescription,poyatos1997complete} of the gates, utilizing a maximum-likelihood estimation approach~\cite{hradil2004ml} to obtain the Choi state corresponding to the operation of each gate~\cite{jamiolkowski1972linear,choi1975completely,jiang2013channel} (see Appendix \ref{app:choi}).  Figure~\ref{fig:choi} depicts the absolute difference between the ideal and simulated Choi matrices of these gates in the qubit subspace. The average process fidelities achieved are $99.6(1)\%$ for SWAP and $99.8(1)\%$ for CZ, with corresponding photon loss probabilities of $8.1(2)\%$ and $2.6(2)\%$, respectively. These metrics promote the potential of such devices in deterministically generating useful graph states for photonic quantum information processing.

\begin{figure}
    \centering
    \includegraphics[width=\linewidth]{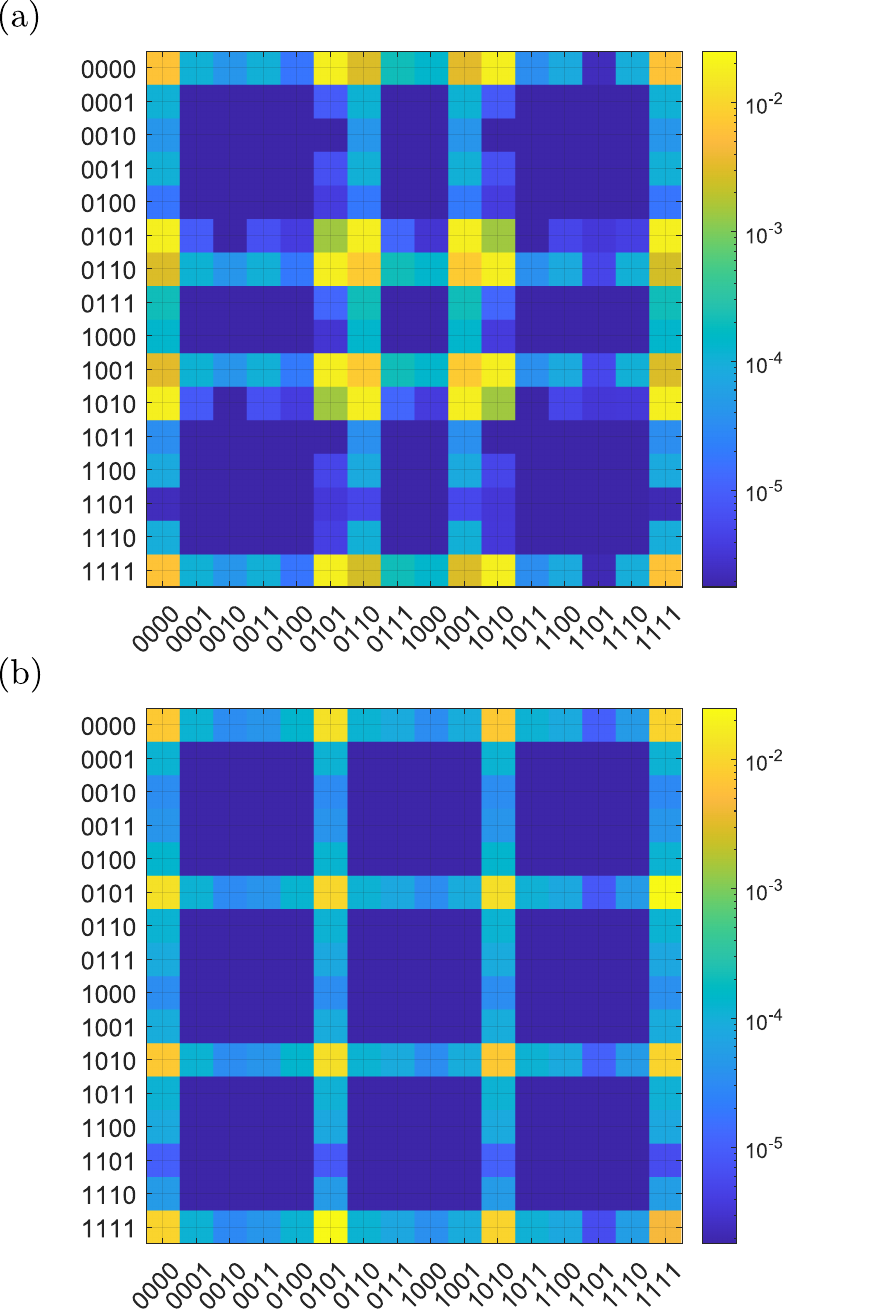}
    \caption{Absolute difference between the ideal and numerically reconstructed Choi matrices for (a) SWAP and (b) CZ in the qubit subspace, based on the $^{87}$Rb implementation of the W-system (see text). The average fidelity for the SWAP and CZ gates is $99.6(1)\%$ and $99.8(1)\%$, respectively.}    
    \label{fig:choi}
\end{figure}

\begin{figure}
    \centering
    \includegraphics[width=\linewidth]{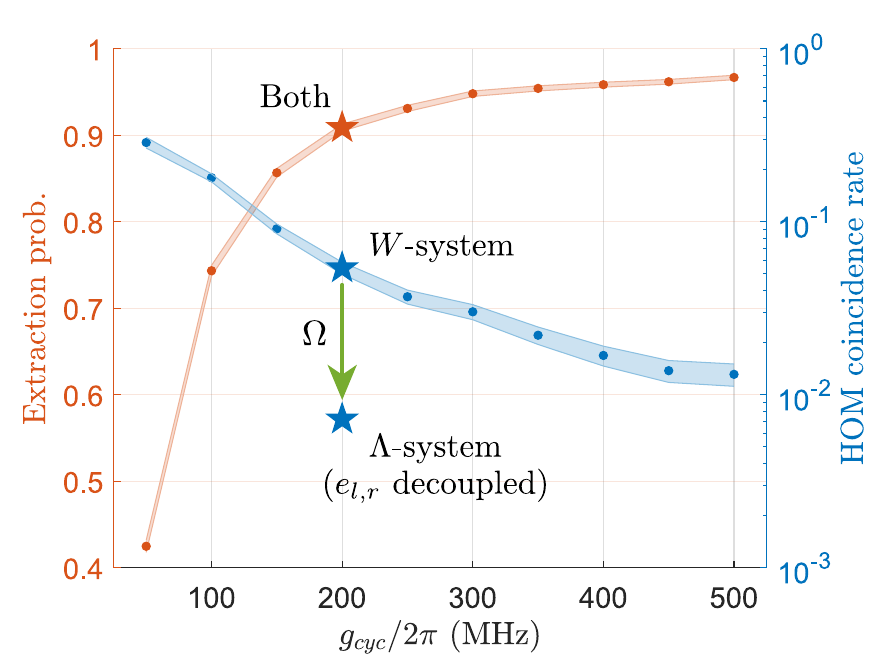}
    \caption{Extraction probability (orange) and HOM coincidence rate in a TQE setup (blue) for different values of $g_{cyc}$ in $^{87}$Rb. Dots and shaded areas indicate the simulated values and uncertainty, respectively. Star markers represent the single-photon source performance for our choice of parameters in W- and $\Lambda$-systems (see text).}
    \label{fig:sps_realistic}
\end{figure}

Furthermore, we investigate the performance of SPRINT-based extraction as a single-photon source. We utilize an incident Gaussian coherent pulse of $\sigma_t=100\text{ns}$ and $\bar{n}=5$, ensuring a negligible vacuum component and maintaining a narrow extracted photon spectrum compared to a pure single photon with $\sigma_t=25\text{ns}$. As the extraction process strongly depends the cooperativity of transitions $\ket{s_{l,r}} \leftrightarrow \ket{e_0}$, we simulate the success probability for different values of $g_{cyc}$ while keeping all other parameters constant, as depicted in Fig.~\ref{fig:sps_realistic}. In the limit of high cooperativity, the extraction probability is only limited by the intrinsic losses of the resonator~\cite{rosenblum2017analysis}.

As mentioned earlier, modally-impure extracted photons can be reliably employed in two-photon linear-optics operations by utilizing a TQE, which relies on a photon number parity measurement in an interferometer involving the transmitted fields from two extraction sources~\cite{aqua2024temporal}. This ensures that the photonic graph states produced in our scheme are compatible with architectures requiring such operations, e.g. fusion-based quantum computation~\cite{bartolucci2023fusion}. However, in practical systems, physical imperfections limit the performance in of the TQE scheme. Notably, photon loss in the transmitted fields, occurring between the two extraction events, whether through spontaneous emission or intrinsic cavity losses, introduces which-path information that degrades the TQE interference and hence hinders the accurate classification of the symmetry of the two-photon state.

Moreover, in the W-system, once a photon is extracted, the remaining photons within the incoming extraction pulse interact with an effective two-level atom, as described in Eq. \ref{eq:evo_2level}. Yet, due to spontaneous atomic emission to free-space modes, scattering of these photons can result in atomic decay to non-interacting states in $\ket{F=2}$. This causes the phase of the transmitted field to flip as it interacts with an empty cavity, similarly to Eq. \ref{eq:condpi}. Consequently, this leads to unwanted detection events at the dark port that can mislead the parity measurement and compromise the operation of the TQE. This effect can be mitigated by postselecting cases where the atoms remains in the $\ket{F=1}$ manifold after the process or by considering only detection events at the dark port that occur between the extraction events \cite{TQEremark}. 

Alternatively, by utilizing the control field with $\Delta_l\!=\!\Delta_r\!=\!0$ to suppress transitions to $e_l$ and $e_r$, photon extraction can be performed in an effective atomic $\Lambda$-type level scheme. In this configuration, following extraction, the atom is driven to the corresponding dark state in $\ket{F=1}$, thereby eliminating the probability of atomic excitation and spontaneous emission to $\ket{F=2}$ states.  Note that the interaction with the first photon of the incoming pulse is the same in both the $\Lambda$- and W-type systems, hence the extraction probabilities in Fig.~\ref{fig:sps_realistic} apply for the $\Lambda$-system as well. 

We simulate the Hong-Ou-Mandel (HOM) interference~\cite{hong1987measurement} between a pair of extracted photons in a TQE setup. We obtain the HOM coincidence rate heralded upon an even number of detection events in the dark port of the TQE interferometer. This rate serves as a measure of the infidelity in two-photon linear-optics gates. As depicted in Fig.~\ref{fig:sps_realistic}, for our choice of parameters, the $\Lambda$-system configuration outperforms the $W$-system, resulting in a coincidence rate of $0.7(1)\%$ and $5.4(1)\%$, respectively. Since the photon loss experienced by the transmitted field is similar in both configurations, the main factor contributing to the difference in the coincidence rate is spontaneous emission to the $\ket{F=2}$ manifold in the W-system. In this configuration, a higher cooperativity of the effective two-level transition lowers both the decay probability to $\ket{F=2}$ and the photon loss rate. This is reflected in Fig.~\ref{fig:sps_realistic}, where the coincidence rate decreases with increasing $g_{cyc}$. In contrast, in the $\Lambda$-system, photon loss in the transmitted field is determined by the resonator quality, characterized by the ratio of $\kappa_e$ to $\kappa_i$. As a result, the coincidence rate is expected to remain constant across different values of $g_{cyc}$.

Our numerical analysis in this section addresses non-ideal photon-atom operations involving a limited-bandwidth atom-cavity system, a complex atomic level scheme, and undesirable loss channels represented by $\kappa_i$ and $\gamma$. However, practical systems may exhibit additional imperfections that introduce errors to these operations, such as atomic motion, parasitic coupling between the modes of the resonator, and polarization mismatch between these modes and the atomic transitions~\cite{rosenblum2017analysis}. Some of these effects can be mitigated through physical techniques, such as the application of a magnetic field and spectral filtering, or by incorporating heralding mechanisms to favor the desired outcome~\cite{chan2022quantum}. We defer to future work a more comprehensive analysis, which explores these aspects in the available parameter space and establishes a connection between imperfect photon-atom operations and a complete characterization of generated graph states, taking into account qubit leakage and correlated errors arising during the generation process.

\section{Discussion}
\label{sec:diss}

We have introduced a scheme for a deterministic, atom-based generation of photonic graph states. Our scheme employs cavity-mediated atomic W-systems capable of both producing single photons through SPRINT-based extraction and executing high-fidelity photon-atom SWAP and CZ gates. These gates maintain near-unit fidelity even with photons in mixed temporal modes, ensuring their compatibility with the extracted photons. By integrating these processes within an interconnected arrangement of W-systems, our scheme offers a deterministic, fast and versatile method for generating arbitrary photonic graph states. A unique feature of our construction is the ability to deterministically stitch together graph states produced by distinct modules, each comprising a few atom-cavity nodes, through entangling operations in a dedicated module. This enables a modular architecture, where individual modules can operate independently to generate and entangle graph states, supporting efficient use of physical resources and simplifying scaling to larger systems. Additionally, by employing the TQE, pairs of extracted photons can be projected into indistinguishable two-photon states, enabling the integration of our generated graph states into linear-optics-based fusion networks.

Moreover, we have explored the implementation of the atomic W-system in $^{87}$Rb under realistic conditions though numerical simulations and evaluated key performance metrics for the photon-atom SWAP and CZ gates, as well as for single-photon extraction and the TQE. These metrics demonstrate the potential of this approach and encourage further research to fully assess the integration of such an interface in practical devices. This includes developing a comprehensive error model that accounts for all potential physical imperfections and identifying the optimal operating conditions to support useful quantum applications.

We note that the atom-cavity nodes used in our scheme can either be integrated within the same device (e.g., a photonic chip) and connected via waveguides, or distributed across separate devices and linked by optical fibers. Naturally, each of these architectures presents distinct advantages and challenges. Integrating all nodes within the same device enables efficient use of hardware, such as sharing a common ultracold atom source for multiple single-atom trapping sites. However, the number of nodes is constrained by physical factors, including the footprint of various photonic components on the chip. On the other hand, linking nodes across separate devices increases scalability, and coupling the nodes to optical fibers is, in any case, necessary for exploiting long-fiber photonic quantum memories. Thus, the optimal architecture for large-scale photonic graph state generation must be a hybrid approach, combining both types of interconnects. The modularity of our design, with a uniform structure for each atom-cavity site, provides flexibility in selecting the most suitable interconnects for various tasks. Moreover, it allows each site to be dynamically reconfigured in real-time as either a photon source or an entangling unit, which could serve as a valuable tool in large-scale systems.

Unlike previous photonic graph state generation schemes, our approach assigns single-photon generation and entanglement operations to separate atom-cavity nodes, referred to as the source and gate units, respectively. While this design inherently requires more atoms, it provides versatility in devising protocols for arbitrary graph state generation. Specifically, the same state can be produced using different protocols with varying resource demands, where a natural trade-off arises between the protocol duration and the number of atoms required. For example, the scheme for generating the 6-ring graph state described in Sec. \ref{sec:graph} requires two steps and involves six atom-cavity nodes: three source and three gate units. Alternatively, the same state can be generated using one source unit extracting six sequential photons, which are then directed to the three gate units. This reduces the number of atoms needed from six to four but extends the generation process to six steps instead of two. This trade-off plays an important role in the design of large-scale modular photonic architectures. In particular, storing photonic qubits in optical fibers allows for the amplification of the entanglement provided by a single module generating graph states repetitively, in a process known as interleaving \cite{bombin2021interleaving}. This capability strongly depends on the generation rate, as photon loss limits the usable fiber length. Therefore, a higher repetition rate allows for greater memory capacity and more efficient utilization of the physical resources. Thus, the versatility of our gate-based approach offers a distinct advantage, allowing for the optimization of the generation rate based on the available hardware.

\begin{acknowledgments}
We acknowledge Daniel Sabsovich and Serge Rosenblum for fruitful discussions, and support from the Israeli Science Foundation, the Binational Science Foundation,
and the Minerva Foundation. B.D. is the Dan Lebas and Roth Sonnewend Professorial Chair of Physics.
\end{acknowledgments}

\appendix
\section{Photon-atom gates with photons in mixed spectral modes}
\label{app:dirtygate}
\subsection{SPRINT with a pure single photon}
\label{app:mou_sprint}
In the configuration that leads to SPRINT, with an incoming pure single photon in mode $\hat{a}_r$ and the atom initialized in $\ket{s_l}$, the state of the system at all times can be expressed as,
\begin{align}
    \ket{\psi(t)} = &\int d\omega C_{s_l}(\omega,t) a^\dagger_r(\omega) \ket{0,s_l}  \\
    & +\int d\omega C_{s_r}(\omega,t) a^\dagger_l(\omega) \ket{0,s_r} + C_{e_0}(t) \ket{0,e_0} \nonumber
\end{align}
with initial conditions $C_{s_r}(\omega,t_0) =  C_{e_0}(t_0) = 0$, and the normalization condition $\int d\omega \abs{C_{s_l}(\omega,t_0)}^2 =1$. Note that states $\ket{e_{l,r}}$ are not involved in the interaction since there is a single excitation in the system. The wavefunction evolves under the time-dependent Schr\"{o}dinger equation arising from the 'modes of the universe' system Hamiltonian (Eq. \ref{eq:Hamiltonian}), leading to the following differential equations,
\begin{align}
    \dot{C}_{s_l}(\omega,t) = g& \frac{\sqrt{\kappa/\pi}}{\kappa + i\omega} e^{i(\omega + \delta_a)t} C_e(t) \label{eq:Cg1} \\
    \dot{C}_{s_r}(\omega,t) = g& \frac{\sqrt{\kappa/\pi}}{\kappa + i\omega} e^{i(\omega + \delta_a)t} C_e(t) \label{eq:Cg2}\\
    \dot{C}_{e_0}(t) = -g& \int d\omega \frac{\sqrt{\kappa/\pi}}{\kappa - i\omega}  e^{-i(\omega + \delta_a)t} \label{eq:Ce} \\
    &\times\Bigl[C_{s_l}(\omega,t) +C_{s_r}(\omega,t)\Bigr]   \nonumber
\end{align}
where we denote $g = g_{r0}=g_{l0}$. 

We first formally integrate Eqs. \ref{eq:Cg1} and \ref{eq:Cg2}, 
\begin{align}
    C_{s_l}(\omega,t) &=  C_{s_l}(\omega,t_0) \label{eq:Cg1ii} \\ &+g \frac{\sqrt{\kappa/\pi}}{\kappa + i\omega} \int_{t_0}^{t} dt' C_{e_0}(t') e^{i(\omega + \delta_a)t'}  \nonumber \\
    C_{s_r}(\omega,t) &=   g \frac{\sqrt{\kappa/\pi}}{\kappa + i\omega} \int_{t_0}^{t} dt' C_{e_0}(t') e^{i(\omega + \delta_a)t'} \label{eq:Cg2ii}  
\end{align}
Substituting Eqs. \ref{eq:Cg1ii} and \ref{eq:Cg2ii} into Eq. \ref{eq:Ce} and performing the second integral over $\omega$,
\begin{align}
    \dot{C}_{e_0}(t) = &-g \int d\omega \frac{\sqrt{\kappa/\pi}}{\kappa - i\omega}  C_{s_l}(\omega,t_0) e^{-i(\omega + \delta_a)t} \label{eq:Cedot} \\
    &-2g^2 \int_{t_0}^{t} dt' C_{e_0}(t') e^{-(\kappa+i\delta_a)(t-t')} \nonumber
\end{align}
Now introduce the Fourier transform,
\begin{equation}
    C_{e_0}(t) = \frac{1}{2\pi}\int d\eta \Tilde{C}_{e_0}(\eta) e^{-i\eta t}
\end{equation}
into Eq. \ref{eq:Cedot},
\begin{align}
    &\frac{1}{2\pi}\int d\eta (-i\eta) \Tilde{C}_{e_0}(\eta) e^{-i\eta t} = \label{eq:Cee} \\ &-g \int d\omega \frac{\sqrt{\kappa/\pi}}{\kappa - i\omega}  C_{s_l}(\omega,t_0) e^{-i(\omega + \delta_a)t} \nonumber \\ &-\frac{g^2}{\pi} \int_{t_0}^{t} dt'  \int d\eta \Tilde{C}_{e_0}(\eta) e^{-i\eta t' -(\kappa+i\delta_a)(t-t')} \nonumber
\end{align}
Fourier transforming Eq. \ref{eq:Cee} (integrating both sides $\int dt e^{i\eta't}$) and formally letting $t_0\!\rightarrow\!-\infty$ allows us to solve for $\Tilde{C}_{e_0}(\eta)$,
\begin{equation}
    \Tilde{C}_{e_0}(\eta) = -2g\sqrt{\kappa \pi} \frac{C_{s_l}(\eta - \delta_a, -\infty)}{2g^2 - i\eta(\kappa - i(\eta - \delta_a))}  \label{eq:Ceta}
\end{equation}
We can subtitute Eq. \ref{eq:Ceta} back into Eqs. \ref{eq:Cg1ii} and \ref{eq:Cg2ii} by first observing that for the final state of the system, at $t\!\rightarrow\!\infty$, the ground state amplitudes are simply given by,
\begin{align}
    C_{s_l}(\omega,\infty) = &C_{s_l}(\omega,-\infty) + g \frac{\sqrt{\kappa/\pi}}{\kappa + i\omega} \Tilde{C}_{e_0}(\omega + \delta_a)  \label{eq:Cg1iii}   \\
    C_{s_r}(\omega,\infty) =  &g \frac{\sqrt{\kappa/\pi}}{\kappa + i\omega} \Tilde{C}_{e_0}(\omega + \delta_a) \label{eq:Cg2iii}  
\end{align}
We can then write,
\begin{align}
    C_{s_l}(\omega,\infty) = \; &C_{s_l}(\omega,-\infty) \Bigl( 1 - S(\omega) \Bigr)\label{eq:Cg1f}   \\
    C_{s_r}(\omega,\infty) =  & -C_{s_l}(\omega,-\infty) S(\omega) \label{eq:Cg2f}  
\end{align}
where,
\begin{align}
    &S(\omega) =  
    \frac{2g^2\kappa}{(\kappa+i\omega)(2g^2 - i(\omega+\delta_a)(\kappa - i\omega))} 
\end{align}

Since the outgoing field in each mode is given by $C_{s_l}(\omega,\infty)$ and $C_{s_l}(\omega,\infty)$ multiplied by the cavity transmission factor $(i\omega + \kappa)/(i\omega - \kappa)$~\cite{gea2011comparative}, the overall state evolution is,
\begin{alignat}{2}
    \label{eq:sprint_evo}
    &\int d\omega f(\omega) &&a^\dagger_r(\omega) \ket{0,s_l} \longrightarrow \\
    &\int d\omega f(\omega) \Bigr[&& J(\omega) a^\dagger_l(\omega) \ket{0,s_r} \nonumber \\ 
    & && + \Bigl(J(\omega) - \frac{\kappa + i\omega}{\kappa - i\omega} \Bigr) a^\dagger_r(\omega) \ket{0,s_l} \Bigl] \nonumber
\end{alignat}
where we denote,
\begin{align}
    f(\omega) = C_{s_l}(\omega,-\infty) \\
    J(\omega) = \frac{\kappa + i\omega}{\kappa - i\omega} S(\omega)
\end{align}

Note the final state remains normalized, as $\abs{\frac{\kappa + i\omega}{\kappa - i\omega}}^2=1$ and $S(\omega)$ maintains,
\begin{equation}
    \abs{S(\omega)}^2 + \abs{1 - S(\omega)}^2 = 1
\end{equation}

\subsection{SPRINT with a photon in a mixed state of spectral modes}
Consider the initial state for SPRINT with an incoming single photon in a mixed state of spectral modes,
\begin{equation}
    \rho_\text{ini} = \iint d\omega_1 d\omega_2 D(\omega_1,\omega_2) a^\dagger_r(\omega_1)\ket{0,s_l}\bra{0,s_l}a_r(\omega_2) 
\end{equation}
We can always choose a different basis in which the density matrix is diagonlized, i.e. written as a statistical mixture of pure single photon states,
\begin{equation}
    \rho_\text{ini} = \sum_i  P_i \ket{\psi_i}\bra{\psi_i}
    \label{eq:eigmod}
\end{equation}
where,
\begin{align}
    \ket{\psi_i} = \int d\omega f_i(\omega) a^\dagger_r(\omega) \ket{0,s_l}
\end{align}
and,
\begin{equation}
    D(\omega_1,\omega_2) = \sum_i P_i f_i(\omega_1)f_i^*(\omega_2)
    \label{eq:Cw1w2}
\end{equation}

Assuming the eigenmodes $f_i(\omega)$ are well-behaved (absolutely integrable), we can evolve each of the pure states separately according to Eq. \ref{eq:sprint_evo}. We can then use Eq. \ref{eq:Cw1w2} to write the final state of the system, 
\begin{widetext}
    \label{eq:eigmod_evo}
    \begin{align}
    \rho_f = \iint d\omega_1 d\omega_2 D(\omega_1,\omega_2) \Bigl[& \Bigl(J(\omega_1) - \frac{\kappa + i\omega_1}{\kappa - i\omega_1} \Bigr)\Bigl(J^*(\omega_2) - \frac{\kappa - i\omega_2}{\kappa + i\omega_2} \Bigr) a^\dagger_r(\omega_1) \ket{0,s_l}\bra{0,s_l} a_r(\omega_2)  \\
    & \Bigl(J(\omega_1) - \frac{\kappa + i\omega_1}{\kappa - i\omega_1} \Bigr)J^*(\omega_2) a^\dagger_r(\omega_1) \ket{0,s_l}\bra{0,s_r} a_l(\omega_2) \nonumber \\
    & J(\omega_1)\Bigl(J^*(\omega_2) - \frac{\kappa - i\omega_2}{\kappa + i\omega_2} \Bigr) a^\dagger_l(\omega_1) \ket{0,s_r}\bra{0,s_l} a_r(\omega_2) \nonumber \\
    & J(\omega_1)J^*(\omega_2)a^\dagger_l(\omega_1) \ket{0,s_r}\bra{0,s_r} a_l(\omega_2) \Bigr] \nonumber
    \end{align}
\end{widetext}

We define the fidelity of SPRINT as the probability of finding the atom in $\ket{s_r}$, and accordingly, the output photon in mode $\hat{a}_l$,
\begin{align}
    \mathcal{F}_{sprint} &= \int d\omega \bra{0,s_r} a_l(\omega) \rho_f a^\dagger_l(\omega) \ket{0,s_r} \\
    &= \int d\omega D(\omega,\omega) \abs{J(\omega)}^2 \nonumber
\end{align}
As evident, SPRINT fidelity depends solely on the power spectrum of the incoming photon.

\subsection{Conditional $\pi$-phase shift with a photon in a mixed state of spectral modes}
\label{app:condpi}
When the control field is activated, effectively setting $g_{r0}=g_{l0}=0$, the W-system is in the CZ operation mode. In this configuration, for an incoming pure single photon in mode $\hat{a}_r$, the state of the system at all times can be expressed as,
\begin{align}
    \ket{\psi(t)} = &\int d\omega C_{s_l}(\omega,t) a^\dagger_r(\omega) \ket{0,s_l} +  \\
    & \int d\omega C_{s_r}(\omega,t) a^\dagger_r(\omega) \ket{0,s_r} + C_{e_r}(t) \ket{0,e_r} \nonumber
\end{align}
Note that $\ket{e_r}$ is the only excited state coupled to by the system Hamiltonian (Eq. \ref{eq:Hamiltonian}) in this setting.

When the atom is prepared in $\ket{s_l}$, it does not interact with the photon and  we can immediately write,
\begin{align}
    C_{s_l}(\omega,\infty) = \; &C_{s_l}(\omega,-\infty) \label{eq:Cg1fcz} 
\end{align}
However, for an atom initialized in $\ket{s_r}$ the incoming photon effectively interacts with a two-level atom and the evolution of the state is given by~\cite{gea2013space}, 
\begin{align}
    C_{s_r}(\omega,\infty) = \; &C_{s_r}(\omega,-\infty) Q(\omega) \label{eq:Cg2fcz} 
\end{align}
where we define,
\begin{align}
    &Q(\omega) =  \frac{\omega + i\kappa}{\omega - i\kappa} \frac{(\omega - i\kappa)(\omega+\delta_a) - g_{rr}^2}{(\omega + i\kappa)(\omega+\delta_a) - g_{rr}^2} \nonumber
\end{align}

Similarly to Eq. \ref{eq:sprint_evo}, the output field in mode $\hat{a}_r$ is given by multiplying the amplitudes in Eq. \ref{eq:Cg1fcz}-\ref{eq:Cg2fcz} by the cavity transmission factor $-(\kappa + i\omega)/(\kappa - i\omega)$. Hence, we can write the evolution of the following states,
\begin{align}
    \int d\omega f(\omega)  a^\dagger_r(\omega) \ket{0,s_l} \rightarrow  -\int d\omega f(\omega) \frac{\kappa + i\omega}{\kappa - i\omega}  a^\dagger_r(\omega) \ket{0,s_l} \\
    \int d\omega f(\omega)  a^\dagger_r(\omega) \ket{0,s_r} \rightarrow  \int d\omega f(\omega) \chi(\omega) a^\dagger_r(\omega) \ket{0,s_r}
\end{align}
where $f(\omega)$ denotes the spectral amplitude profile of the incoming photon and we define,
\begin{align}
    &\chi(\omega) =  \frac{(\omega - i\kappa)(\omega+\delta_a) - g_{rr}^2}{(\omega + i\kappa)(\omega+\delta_a) - g_{rr}^2} \nonumber
\end{align}
which maintains $\abs{\chi(\omega)}^2 = 1$.

In order to examine the conditional phase shift with an incoming photon in a mixed state of spectral modes, we consider an atom initialized in an equal superposition state, $(\ket{s_l}+\ket{s_r})/2$,
\begin{align}
    \rho_\text{ini} = &\frac{1}{2}\iint d\omega_1 d\omega_2 D(\omega_1,\omega_2) a^\dagger_r(\omega_1)\ket{0}\bra{0}a_r(\omega_2) \\
    &\otimes \Bigl( \ket{s_l}\bra{s_l}+\ket{s_l}\bra{s_r}+\ket{s_r}\bra{s_l}+\ket{s_r}\bra{s_r} \Bigr) \nonumber
\end{align}
Using the same decomposition as in Eq. \ref{eq:eigmod}-\ref{eq:Cw1w2} and the evolution in Eq. \ref{eq:Cg1fcz}-\ref{eq:Cg2fcz}, the final state of the system is given by,
\begin{alignat}{2}
    \label{eq:rhofcz}
    \rho_f = \frac{1}{2}&\iint && d\omega_1 d\omega_2 D(\omega_1,\omega_2) a^\dagger_r(\omega_1)\ket{0}\bra{0} a_r(\omega_2) \\ 
    &\otimes \Bigl[&& \ket{s_l}\bra{s_l} -\chi^*(\omega_2) \frac{\kappa + i\omega_1}{\kappa - i\omega_1} \ket{s_l}\bra{s_r}  \nonumber \\
     & &&  -\chi(\omega_1)\frac{\kappa - i\omega_2}{\kappa + i\omega_2} \ket{s_r}\bra{s_l} + \ket{s_r}\bra{s_r}  \Bigr] \nonumber
\end{alignat}

Ideally, the photon induces a relative $\pi$ phase shift and atom ends up in the orthogonal superposition state, $(\ket{s_l}-\ket{s_r})/2$. Therefore, the fidelity of the process is given by,
\begin{align}
    \mathcal{F}_{cp} &= \int d\omega \frac{\bra{s_l}-\bra{s_r}}{\sqrt{2}}\bra{0} a_r(\omega) \rho_f a^\dagger_r(\omega) \ket{0}\frac{\ket{s_l}-\ket{s_r}}{\sqrt{2}} \nonumber \\
    &= \frac{1}{2}\int d\omega D(\omega,\omega) \Bigr(1 + \Re (\chi(\omega)\frac{\kappa - i\omega}{\kappa + i\omega})\Bigl) 
\end{align}
which only depends on the power spectrum of the incoming photon.

\section{Theoretical model for numerical simulations}
\label{app:simulation}

\begin{figure}
    \centering
    \includegraphics[width=\linewidth]{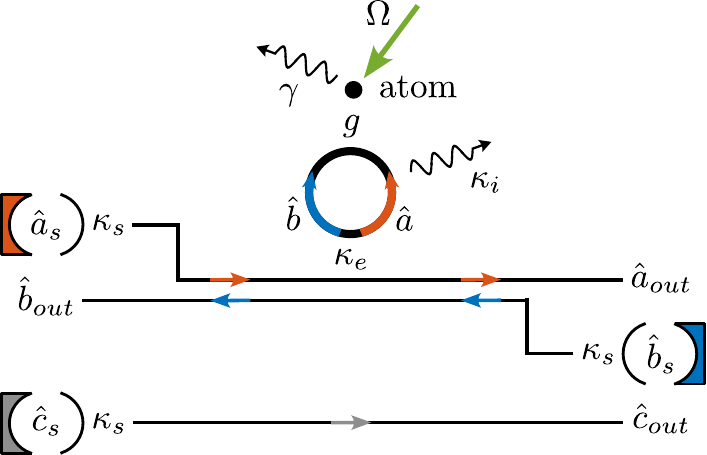}
    \caption{Schematic description of the atom-cavity system within the framework of cascaded open quantum systems (see text).}
    \label{fig:inputoutput}
\end{figure}

In this paper, we derive analytical expressions based on the modes of the universe Hamiltonian in Eq. \ref{eq:Hamiltonian}, which describes the cavity-mediated interaction of quantum fields $\hat{a}_l$ and $\hat{a}_r$ with a single atom. To complement these results and extend our analysis beyond ideal photon-atom operations, we adopt a description our system using the theory of cascaded open quantum systems~\cite{carmichael1993quantum}, as depicted in Fig.~\ref{fig:inputoutput}; one-sided virtual cavities $\hat{a}_s$ and $\hat{b}_s$ decay with rate $\kappa_s$ into orthogonal modes of a waveguide, which couple with rate $\kappa_e$ to modes $\hat{a}$ and $\hat{b}$ of the system resonator, respectively. The optical field in the resonator interacts with the atom with coupling rate $g$ and eventually leaks back into the waveguide, yielding output operators~\cite{gardiner1985input},
\begin{align}
    &a_{out} = \sqrt{2\kappa_s}a_s + \sqrt{2\kappa_e}a \label{eq:output_a}\\
    &b_{out} = \sqrt{2\kappa_s}b_s + \sqrt{2\kappa_e}b \label{eq:output_b}
\end{align}
An additional virtual cavity decays with the same rate $\kappa_s$ into a different waveguide that does not interact with the resonator, producing the output mode,
\begin{align}
    &c_{out} = \sqrt{2\kappa_s}c_s \label{eq:output_c}
\end{align}

We employ the quantum trajectories approach~\cite{carmichael1993open}, also known as the Monte-Carlo wavefunction method~\cite{dalibard1992wave,molmer1993monte}, to simulate the evolution of stochastic wavefunctions according to the Schr\"{o}dinger equation with a non-Hermitian Hamiltonian and quantum jump operators. Following~\cite{carmichael1993quantum,rosenblum2017analysis}, the state of our system evolves with a non-Hermitian Hamiltonian given by,
\begin{align}
    \frac{\mathcal{H}}{\hbar}=&-2i\sqrt{\kappa_s\kappa_e}(a_sa^\dagger+b_sb^\dagger) -i\kappa_s(a_s^\dagger a_s + b_s^\dagger b_s + c_s^\dagger c_s) \nonumber \\
    &+\sum_{j,k}\Bigl(g^a_{jk}a^\dagger\ket{s_j}\bra{e_k} + g^b_{jk}b^\dagger\ket{s_j}\bra{e_k} + \text{h.c}\Bigr) \nonumber \\
    &+ (\kappa + i\delta_C)(a^\dagger a + b^\dagger b) + \sum_k(\gamma + i\delta_k)\ket{e_k}\bra{e_k} \nonumber \\
    & + \sum_j\Delta_j\ket{s_j}\bra{s_j}
\end{align}
Here, states $\ket{s_j}$ ($\ket{e_k}$) represent ground (excited) states, while $g^a_{jk}$ and $g^b_{jk}$ indicate the coupling strength of modes $\hat{a}$ and $\hat{b}$ to atomic transition $\ket{s_j}\leftrightarrow\bra{e_k}$, respectively. Each atomic transition experiences a free-space decay rate denoted by $\gamma_{jk}$, with the total decay rate of each excited state being $\gamma=\sum_j{\gamma_{jk}}$. In our discussion of the $^{87}$Rb implementation in Sec. \ref{sec:feas}, these parameters are determined by atomic selection rules, setting $g^{a/b}_{jk}=\gamma_{jk}=0$ for forbidden transitions, and by appropriate branching ratios that establish their relative magnitudes. The total decay rate of the resonator is given by $\kappa = \kappa_e + \kappa_i$, where $\kappa_i$ denotes the intrinsic loss rate. $\delta_C$ and $\delta_k$ denote the detuning of the resonator and the atomic transitions relative to the frequency of the virtual cavities, respectively. $\Delta_j$ denotes the frequency shift of each of the ground states relative to a lowest energy level. 

The quantum jump operators in our system include the output modes of the waveguides as described in Eq. \ref{eq:output_a}-\ref{eq:output_c}, along with additional operators accounting for the coupling of the atomic transitions and the modes of the resonator to the environment,
\begin{align}
    &L_{jk} = \sqrt{2\gamma_{jk}}\ket{s_j}\bra{e_k} \\
    &L_{a} = \sqrt{2\kappa_i}a \\
    &L_{b} = \sqrt{2\kappa_i}b 
\end{align}
For each wavefunction, the Monte-Carlo procedure yields photon detection events induced by the jump operators and the corresponding final state of the atom. Measurement probabilities are then calculated by ensemble averages over the wavefunctions. The  uncertainty in these values scales inversely with the square root of the number of simulated wavefunctions and is represented by the error bars in Fig. \ref{fig:sps}-\ref{fig:dirtygates} and shaded area in Fig. \ref{fig:sps_realistic}.

Our treatment of the control field differs based on whether we are analyzing the ideal scenario, in which we simply set $g_{jk}=0$ for the relevant transitions, or the implementation in $^{87}$Rb, where we directly incorporate an interaction Hamiltonian with a higher excited state manifold,
\begin{equation}
\mathcal{H}_\text{control} = \sum_{k,l} \Omega_{kl} \ket{e_k}\bra{e'_l} e^{-i\Delta_{kl}t} + \text{h.c.}
\end{equation}
where $\ket{e'_l}$ denotes the higher excited states, while $\Omega_{kl}$ and $\Delta_{kl}$ represent the Rabi frequency and relative detuning of the external field for each transition, respectively. 

\begin{figure*}[t]
    \centering
    \includegraphics[width=0.95\linewidth]{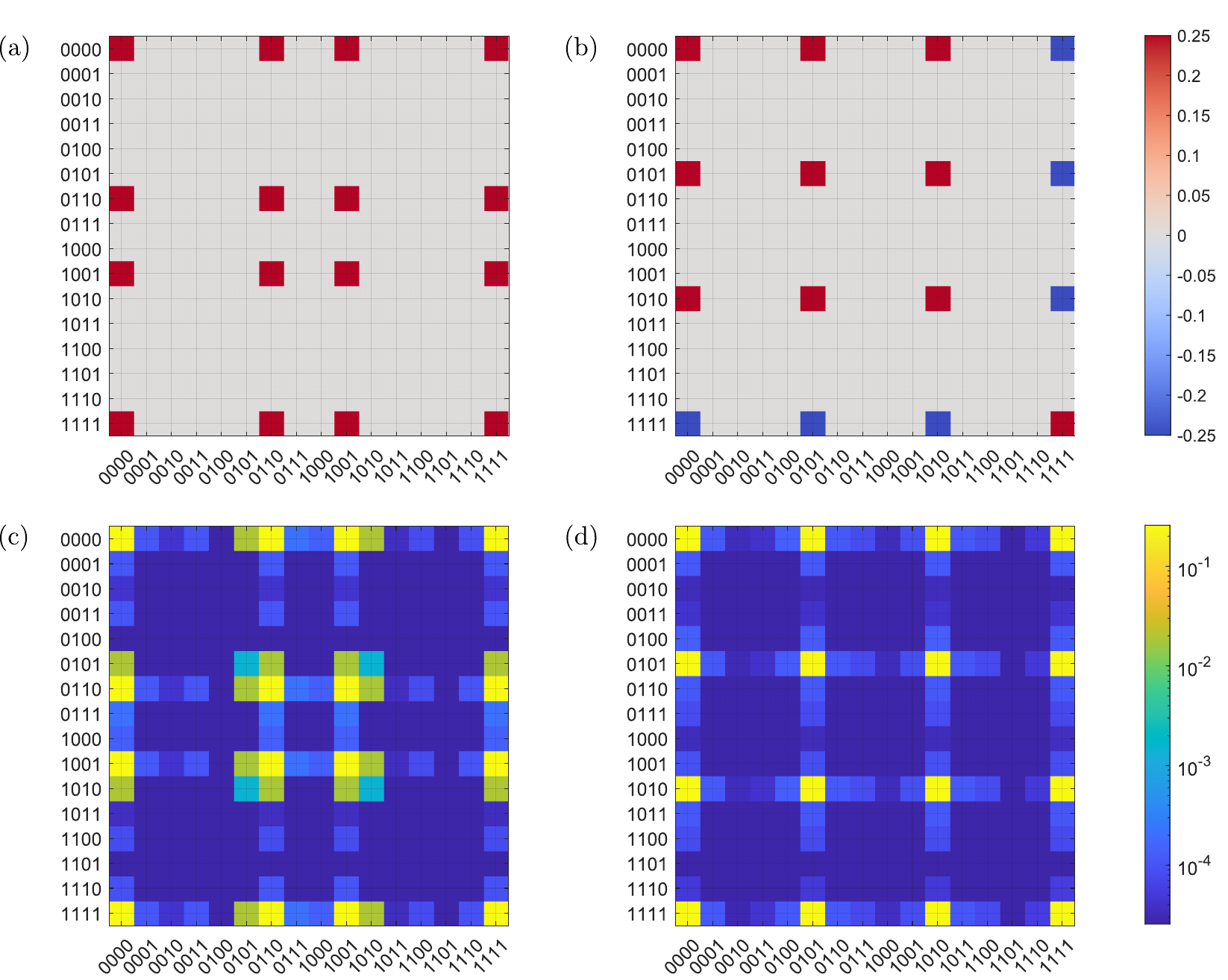}
    \caption{(a)-(b) Choi matrices for the ideal SWAP and CZ gates, respectively. (c)-(d) Reconstructed Choi matrices (absolute values) obtained from simulated process tomography for the photon-atom (c) SWAP and (d) CZ gates in the qubit subspace, based on the $^{87}$Rb implementation of the W-system described in Sec.~\ref{sec:feas}.}    
    \label{fig:choi_appendix}
\end{figure*}

When examining the performance of photon-atom gates with pure photons, we initialize the stochastic wavefunctions with the atom in its qubit subspace and the virtual cavities containing a single excitation; for the SWAP (CZ) gate, the photonic qubit is encoded in a single photon emitted from $\hat{a}_s$ and $\hat{b}_s$ ($\hat{a}_s$ and $\hat{c}_s$). The temporal profile of the photon emitted by the virtual cavity can be controlled by incorporating a time-dependent $\kappa_s$, as detailed in~\cite{kiilerich2019input}. The measurement axis of the output photonic qubit is determined by the definition of the output operators in Eq. \ref{eq:output_a}-\ref{eq:output_c}. Therefore, adjusting the measurement axis involves redefining the output operators as orthogonal superpositions of $\hat{a}_{out}$ and $\hat{b}_{out}$ for SWAP, or $\hat{a}_{out}$ and $\hat{c}_{out}$ for CZ.

For single-photon extraction, the atom is initialized in one of its ground states, and the incident coherent state is enabled by substituting the virtual cavity in the relevant mode with a time-dependent classical field. We employ two copies of this system to evaluate the purity of extracted photons through HOM interference. This involves defining new output modes by coherently mixing the original modes carrying the extracted photons, essentially implementing a beam splitter. Additionally, the effect of the TQE on the HOM coincidence rate is simulated by similarly mixing the output modes carrying the transmitted fields of both extraction sources. Furthermore, to simulate photon-atom gates with extracted photons, we consider two cascaded systems. In this setup, the output of the source system, designed for single-photon extraction, is fed into another system configured to execute photon-atom gates.

\section{The Choi matrix}
\label{app:choi}

A quantum process $\mathcal{E}$ (also referred to as a quantum channel) is generally defined as a linear, completely positive, and trace preserving map from operators on a Hilbert space $\mathcal{H}_A$ to operators on another Hilbert space $\mathcal{H}_B$. Specifically, it transforms an input state $\rho_{\text{in}} \in \mathcal{H}_A$ into an output state $\rho_{\text{out}} = \mathcal{E}(\rho_{\text{in}}) \in \mathcal{H}_B$. The Choi–Jamiołkowski isomorphism establishes a correspondence between such maps and positive semidefinite operators $E$ on the composite Hilbert space $\mathcal{H}_A \otimes \mathcal{H}_B$. This operator can be interpreted as a bipartite quantum state, making the isomorphism also known as the channel-state duality \cite{jiang2013channel}. Hence, the operator $E$ is commonly referred to as the Choi state or Choi matrix.

Consider the maximally entangled state in $\mathcal{H}_A \otimes \mathcal{H}_{A'}$, where $A'$ represent an ancillary system with the same dimensionality as $A$,
\begin{align}
    \ket{\Phi} = \sum_j\ket{j}_A\otimes\ket{j}_{A'}
\end{align}
where $\{\ket{j}\}$ forms a basis of the Hilbert space. The Choi matrix is defined by applying the map $\mathcal{E}$ to the ancillary part of $\ket{\Phi}$,
\begin{align}
    E = \mathcal{I}_A\otimes\mathcal{E}_{A'}(\ket{\Phi}\bra{\Phi}) 
\end{align}
It can then be shown that the operation of $\mathcal{E}$ on an input state $\rho_{\text{in}}$ is given by,
\begin{align}
    \rho_{\text{out}} &= \mathcal{E}(\rho_{\text{in}})= Tr_{\mathcal{H}_A}(E\rho_{\text{in}}^T\otimes I)
\end{align}
In the context of our graph state generation scheme, Fig.~\ref{fig:choi_appendix}(a)-(b) depicts the Choi matrices for the ideal SWAP and CZ gates, where $\{\ket{j}\}$ corresponds to the two-qubit computational basis.

For completeness, we make the connection between the Choi matrix and the more commonly used $\chi$-matrix. The latter describes the operation of the quantum process $\mathcal{E}$ as follows,
\begin{align}
     \mathcal{E}(\rho) = \sum_{i,j} \chi_{ij}P_i \rho P_j^\dagger
\end{align}
where $\{P_i\}$ and $\{P_j\}$ represent basis in the space of density operators, typically chosen to be the Pauli basis for a multi-qubit Hilbert space.
It can be shown that \cite{jong2019fault},
\begin{align}
    \chi_{ij}= \bra{\bra{P_i}} E \ket{\ket{P_j}}
\end{align}
where $\ket{\ket{P_i}}$ are defined as,
\begin{align}
    \ket{\ket{P_i}} = I\otimes P_i \ket{\Phi}
\end{align}

To obtain the Choi matrices for the photon-atom SWAP and CZ gates considered in the main text, we perform quantum process tomography. Using the numerical simulation described in Appendix \ref{app:simulation}, we evolve a set of input quantum states, which forms a basis in the two-qubit space, and calculate the probabilities of a complete set of projective measurements acting on the output states. To reconstruct the quantum process from these probabilities, direct linear inversion \cite{chuang1997prescription,nielsen2010quantum} can be applied. However, due to the finite precision of the simulation output, this method often yields unphysical results. Maximum-likelihood reconstruction of the Choi matrix provides a more robust approach \cite{hradil2004ml}. Using this method, we obtained the Choi matrices for the photon-atom SWAP and CZ gates shown in Fig.~\ref{fig:choi_appendix}(c)-(d). Figure~\ref{fig:choi} in the main text depicts the absolute difference between these matrices and their ideal counterparts. The average process fidelity is simply calculated as the state overlap between the ideal and the reconstructed Choi matrices.

\begin{figure}
    \centering
    \vspace{0.5cm}
    \includegraphics[width=\linewidth]{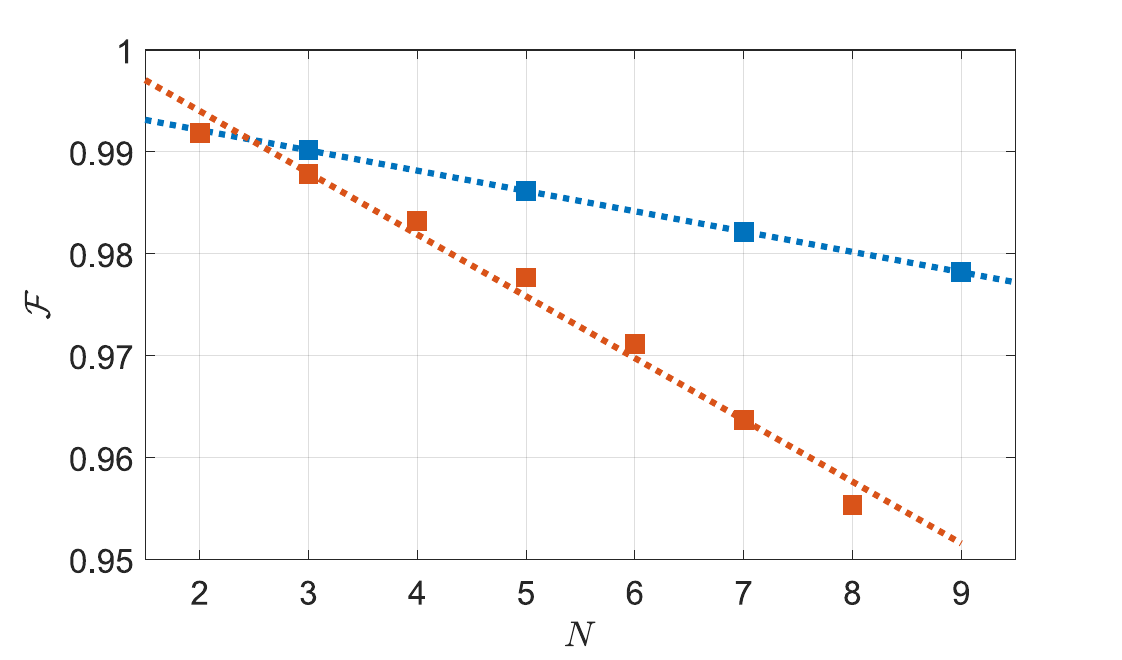}
    \caption{Fidelity of an $N$-photon graph state generated using unidirectional CZ gates (orange) and DCZ gates (blue), with infidelities per added photon of $0.6(1)\%$ and  $0.2(1)\%$, respectively. }    
    \label{fig:fid_scale}
\end{figure}

Using the simulated Choi matrices of the photon-atom CZ and SWAP in the qubit subspace, we can construct multi-qubit operations for the generation of photonic graph states. As an example, we calculate the fidelity of an $N$-photon star graph state, which is equivalent to an $N$-photon GHZ state under local unitaries. We begin by initializing the atomic qubit in a $\ket{+}$ state using SWAP, followed by $(N-1)$ CZ gates between the atomic qubit and $(N-1)$ photonic qubits initialized in $\ket{+}$. This produces a star graph state with the atom as the central qubit. Finally, an additional photon is used to map the atomic qubit to a photonic qubit via another SWAP gate, resulting in an $N$-photon star graph state. The fidelity of this state can be calculated as the overlap with the ideal state. Alternatively, for odd N, the symmetry of the W-system can be harnessed by applying $\frac{N-1}{2}$ DCZ gates. Since the DCZ gate involves both left- and right-propagating photons, it effectively prevents the accumulation of error that occurs in consecutive application of CZ gates in one direction. As shown in Fig.~\ref{fig:fid_scale}, this results in a lower infidelity per added photon to the graph; $0.2(1)\%$ for the DCZ gates compared to $0.6(1)\%$ with unidirectional CZ gates.

\bibliography{refs.bib}

\begin{thebibliography}{127}
\bibitem [{TQE()}]{TQEremark}%
  \BibitemOpen
  \href@noop {} {}\bibinfo {note} {In the ideal TQE scheme, dark port detections can only occur between the two extraction events, and their parity determines the symmetry of the two-photon state. However, atomic decay to the $\ket{F=2}$ manifold through spontaneous emission can result in additional detections in the dark port, even after the extraction events have concluded. To mitigate this effect, we can consider only those detections that fall within the time window between the two extraction events. This would require using the detection times of the extraction photons, following their intended two-photon linear-optics gate, to identify which dark port counts should be included.}\BibitemShut {Stop}%
\end{thebibliography}

\end{document}